\documentclass[aps,amsmath,amssymb,twocolumn]{revtex4-1}

\usepackage{graphicx}
\usepackage{dcolumn}
\usepackage{bm}
\usepackage{color}
\usepackage[normalem]{ulem}
\usepackage{amsmath}

\begin{document}

\title{Interplay between Yu-Shiba-Rusinov states and multiple Andreev reflections}

\author{A. Villas$^{1,\dagger}$}

\author{R.~L. Klees$^{2,\dagger}$}

\author{H. Huang$^{3}$}

\author{C.~R. Ast$^{3}$}

\author{G. Rastelli$^{2,4}$}

\author{W. Belzig$^{2}$}

\author{J.~C. Cuevas$^{1}$}

\affiliation{$^1$Departamento de F\'{\i}sica Te\'orica de la Materia Condensada
and Condensed Matter Physics Center (IFIMAC), Universidad Aut\'onoma de Madrid,
E-28049 Madrid, Spain}

\affiliation{$^2$Fachbereich Physik, Universit{\"a}t Konstanz, D-78457 Konstanz, Germany}

\affiliation{$^3$Max-Planck-Institut f\"ur Festk\"orperforschung, Heisenbergstra{\ss}e 1, 
70569 Stuttgart, Germany}

\affiliation{$^4$Zukunftskolleg, Universit{\"a}t Konstanz, D-78457 Konstanz, Germany}

\date{\today}

\begin{abstract}
Motivated by recent scanning tunneling microscopy experiments on single magnetic 
impurities on superconducting surfaces, we present here a comprehensive theoretical
study of the interplay between Yu-Shiba-Rusinov bound states and (multiple) Andreev
reflections. Our theory is based on a combination of an Anderson model with broken spin
degeneracy and nonequilibrium Green's function techniques that allows us to describe the 
electronic transport through a magnetic impurity coupled to superconducting leads
for arbitrary junction transparency. Using this combination we are able to elucidate
the different tunneling processes that give a significant contribution to the subgap 
transport. In particular, we predict the occurrence of a large variety of Andreev 
reflections mediated by Yu-Shiba-Rusinov bound states that clearly differ from the 
standard Andreev processes in non-magnetic systems. Moreover, we provide concrete 
guidelines on how to experimentally identify the subgap features originating from these
tunneling events. Overall, our work provides new insight into the role of the spin 
degree of freedom in Andreev transport physics.
\end{abstract}

\maketitle

\section{Introduction}

The competition between magnetism and superconductivity is one of the most fundamental
problems in condensed matter physics. The ability of the scanning tunneling microscope
(STM) to manipulate individual magnetic atoms and molecules has enabled to study
this competition at the atomic scale. One of the most interesting manifestations of 
the interplay between these two antagonistic phases of matter is the appearance of 
the so-called Yu-Shiba-Rusinov (YSR) bound states in the spectrum of a single magnetic
impurity coupled to a superconductor or embedded in a superconducting matrix
\cite{Yu1965,Shiba1968,Rusinov1969}. Numerous STM-based experiments on single magnetic
impurities on surfaces of conventional superconductors have reported the
observation of these superconducting bound states \cite{Yazdani1997,Ji2008,Franke2011,
Menard2015,Ruby2015,Hatter2015,Ruby2016,Randeria2016,Choi2017,Cornils2017,Hatter2017,
Farinacci2018,Brand2018,Malavolti2018,Kezilebieke2019,Senkpiel2019,Etzkorn2018,
Schneider2019,Liebhaber2020,Huang2019a,Huang2019b}, for a recent review see 
Ref.~\cite{Heinrich2018}. Those experiments have elucidated many basic aspects of the 
YSR states such as, for instance, the nature of the many-body ground state 
\cite{Franke2011,Hatter2015,Hatter2017}, the spatial extension of the YSR wave functions 
\cite{Menard2015,Ruby2016,Choi2017}, the spin signature \cite{Cornils2017}, or the role 
of key energy scales like the exchange energy \cite{Farinacci2018} or the impurity-substrate 
coupling \cite{Huang2019a}. Moreover, part of the interest in the YSR states lies in 
the fact that they can serve as building blocks to create Majorana states in designer 
structures such as chains of magnetic impurities 
\cite{Nadj-Perge2014,Ruby2015b,Kezilebieke2018,Ruby2017,Ruby2018}.

Often the STM experiments probing single magnetic impurities on superconducting
substrates are done using a superconducting tip to enhance the energy resolution.
More importantly for this work, the use of superconducting tips enables to study a 
variety of tunneling processes, most notably multiple Andreev reflections (MARs), whose
relative importance depends on the junction transmission. Let us remind that in a 
junction between a normal metal and a superconductor, an Andreev reflection consists
in a tunneling process in which an electron coming from the normal metal is reflected 
as a hole of opposite spin transferring a Cooper pair into the superconductor. In the 
absence of in-gap bound states, this process dominates the subgap transport. In the 
case of a junction with two superconducting electrodes, one can 
additionally have MARs in which quasiparticles undergo a cascade of Andreev
reflections that give rise to a very complex subgap structure in the current-voltage 
characteristics. The microscopic theory of MARs, which was developed in the mid-1990s
\cite{Averin1995,Cuevas1996}, was actually quantitatively confirmed in the context of 
superconducting atomic-size contacts with the help of break-junction techniques 
and the STM \cite{Scheer1997,Scheer1998}. In recent years, several STM-based experiments in 
the context of magnetic impurities on superconducting surfaces have revealed signatures
of the interplay between YSR bound states and Andreev reflections \cite{Ruby2015,
Randeria2016,Farinacci2018,Brand2018}, which demonstrates that this type of system is
ideal to study the role of the spin degree of freedom in these tunneling events. 

Some of the aspects of the interplay between YSR states and Andreev reflections
have already been studied theoretically. For instance, in Ref.~\cite{Andersen2011},
and motivated by experiments in the context of quantum dots, a theoretical study of 
the nonlinear cotunneling current through a spinful quantum dot contacted by two 
superconducting leads was presented. This study concluded that while the subgap
transport is dominated by MARs in the limit of symmetric couplings to the 
superconductors, it is determined by the quasiparticle tunneling into spin-induced
YSR states in the strongly asymmetric case (of relevance for our work). On the other 
hand, Ruby \emph{et al.}\ \cite{Ruby2015} analyzed in the case of a superconducting 
tip, both theoretically and experimentally, the competition between the tunneling of 
single quasiparticles and a resonant Andreev reflection as a function of the junction 
transmission. In Ref.~\cite{Brand2018}, the crossover between the tunnel regime (low 
junction transparency) and the contact regime (high transparency) was theoretically analyzed 
in connection with the experiments reported in that work with a normal tip. In 
spite of these interesting works, there is still no systematic theory studying the 
interplay between YSR states and (multiple) Andreev reflections which identifies all the 
relevant tunneling events that can occur in these impurity systems and that could serve 
as a guide for the experimentalists to look for novels features in the subgap transport. 
The goal of this work is to fill this theoretical gap. For this purpose, we present here 
a theory of the interplay between YSR states and MARs in junctions where single magnetic
impurities are coupled to superconducting leads, with special emphasis on STM-based 
experiments. Our theory is based on a combination of a mean-field Anderson model with 
broken spin degeneracy to describe magnetic impurities and nonequilibrium Green's function
techniques to compute the electronic transport properties. Using this combination we
elucidate the complete set of relevant tunneling processes that can occur in these 
systems and provide precise guidelines to experimentally identify the signatures of 
those processes.

The rest of the manuscript is organized as follows. In section \ref{sec-theory} 
we describe the system under study and present the theoretical tools that
we use to compute the electronic transport properties of magnetic impurities coupled
to superconducting leads. In section \ref{sec-NS} we discuss the results for the case
where only one of the leads is superconducting to study in detail the competition between
single-quasiparticle tunneling and the simplest Andreev reflection. Then, in section
\ref{sec-SS} we present a detailed study of the subgap transport in the case of a magnetic
impurity coupled to a superconducting substrate and a superconducting tip, which is the
central goal of this work. Finally, in section \ref{sec-conclusions} we discuss the new 
lines of research that this work opens and we summarize our main conclusions. 
 
\section{Systems under study and theoretical approach} \label{sec-theory}

\begin{figure}[t]
\includegraphics[width=0.6\columnwidth,clip]{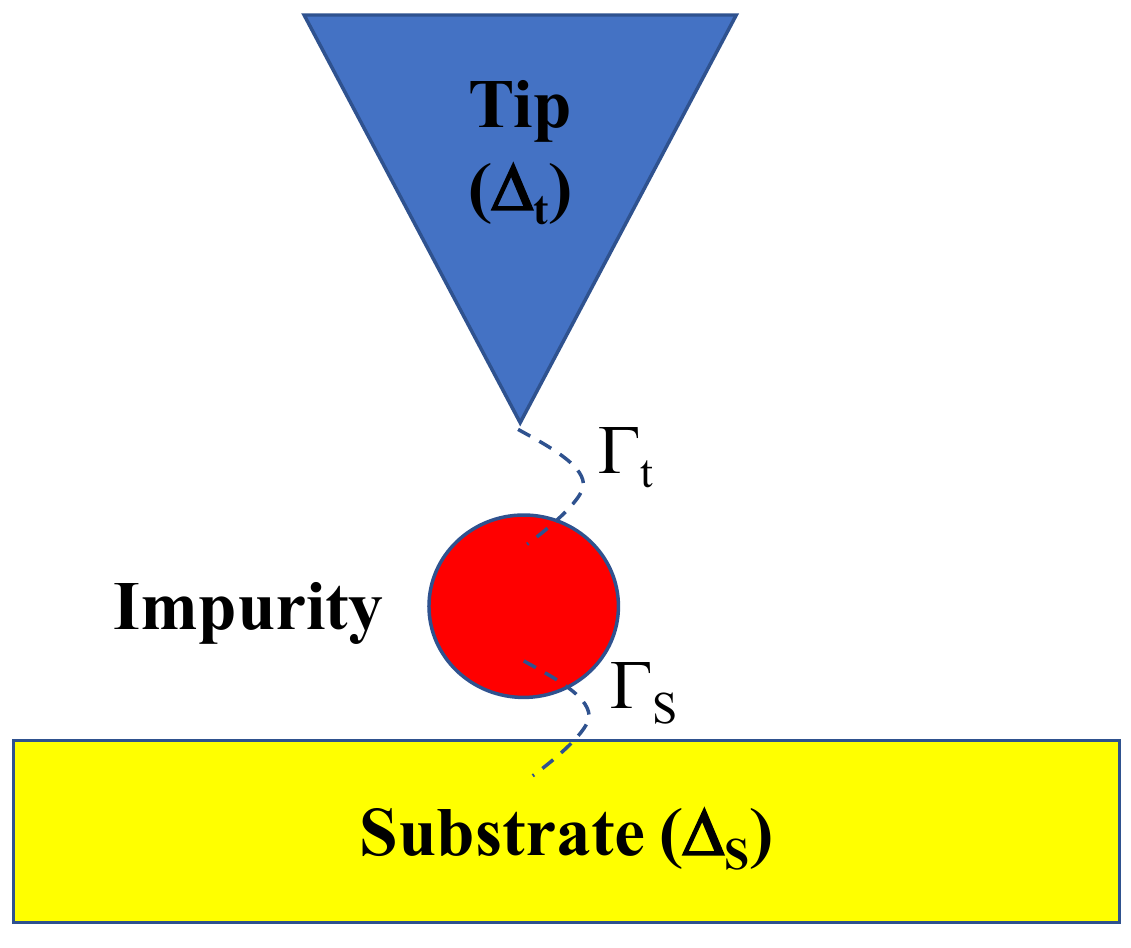}
\caption{Schematic representation of the system under study. A magnetic impurity is
coupled to a superconducting substrate and to an STM tip that can also be superconducting.
The tunneling rates $\Gamma_{\rm t}$ and $\Gamma_{\rm S}$ measure the strength of the 
coupling of the impurity to the tip and substrate, respectively, while $\Delta_{\rm t}$ 
and $\Delta_{\rm S}$ are the corresponding superconducting gaps.}
\label{fig-system}
\end{figure}

Our goal in this work is to compute the current-voltage characteristics in a system in
which a magnetic impurity is coupled to superconducting leads with special emphasis in
the interplay between YSR bound states and MARs. As shown schematically in 
Fig.~\ref{fig-system}, we shall focus on the analysis of the experimentally relevant
situation in which a magnetic impurity (an atom or a molecule) is coupled to a 
superconducting substrate (S) and to an STM tip (t), which can also be superconducting.
This section is devoted to the description of the theoretical tools used to tackle
this problem.

\subsection{The Anderson model}

The Anderson model used here to describe the impurity coupled to superconducting leads
is summarized in the Hamiltonian
\begin{equation}
\label{eq-total-H}
H = H_{\rm t} + H_{\rm S} + H_{\rm i} + H_\mathrm{hopping} .
\end{equation}
Here, $H_j$, with $j={\rm t,S}$, is the BCS Hamiltonian of the lead $j$ given by
\begin{multline}
H_j = \sum_{{\boldsymbol k} \sigma} \xi_{{\boldsymbol k}j} 
c^{\dagger}_{{\boldsymbol k}j\sigma} c_{{\boldsymbol k}j\sigma} \\
+ \sum_{\boldsymbol k} \left( \Delta_j e^{i\varphi_j} 
c^{\dagger}_{{\boldsymbol k}j\uparrow} c^{\dagger}_{{-\boldsymbol k}j\downarrow} 
+ \Delta_j e^{-i\varphi_j} c_{{-\boldsymbol k}j\downarrow} c_{{\boldsymbol k}j\uparrow}
\right) ,
\end{multline}
where $c^{\dagger}_{{\boldsymbol k}j\sigma}$ and $c_{{\boldsymbol k}j\sigma}$ are the creation 
and annihilation operators, respectively, of an electron of momentum ${\boldsymbol k}$, 
energy $\xi_{{\boldsymbol k}j}$, and spin $\sigma=\uparrow, \downarrow$ in lead $j$, 
$\Delta_j$ is the superconducting gap, and $\varphi_j$ is the corresponding superconducting 
phase. On the other hand, $H_{\rm i}$ is the Hamiltonian of the magnetic impurity, which 
in our case is given by
\begin{equation}
H_{\rm i}	= U(n_{\uparrow} + n_{\downarrow}) + J (n_{\uparrow} - n_{\downarrow}) ,
\end{equation}
where $n_{\sigma} = d^{\dagger}_{\sigma} d_{\sigma}$ is the occupation number operator on
the impurity, $U$ is the on-site energy (not to confuse with the Coulomb energy), and 
$J$ is the exchange energy that breaks the spin degeneracy on the impurity. Finally, 
$H_\mathrm{hopping}$ describes the coupling between the magnetic impurity and the leads 
that adopts the form
\begin{equation}
H_{\rm hopping} = \sum_{{\boldsymbol k},j,\sigma} t_j \left(d^{\dagger}_{\sigma}
c_{{\boldsymbol k}j\sigma} + c^{\dagger}_{{\boldsymbol k}j\sigma} d_{\sigma} 
\right) ,
\end{equation}
where $t_j$ describes the tunneling coupling between the impurity and the lead 
$j=\mathrm{t,S}$ and it is chosen to be real.

The Anderson model used here has origin in a mean-field approximation and it provides a
convenient way to describe a quantum magnetic impurity \cite{note1}. In particular, it has
been successfully employed in the past to describe the observation of Andreev bound states 
in quantum dots coupled to superconducting leads, see e.g.\ \cite{Pillet2010}, and it has
been shown to reproduce many of the salient features of the superconducting bound states 
predicted by more sophisticated many-body approaches \cite{Martin-Rodero2011,Martin-Rodero2012}.
Moreover, it has been shown very recently that this model provides a convenient starting
point to illustrate the fundamental role played in the YSR states by the hybridization of 
the impurity with the substrate \cite{Huang2019a}. Finally, the Anderson model (beyond the
mean-field approximation) is ideally suited for studying the role of electronic correlations
in this problem, see e.g.\ Refs.~\cite{Martin-Rodero2011,Martin-Rodero2012,Zitko2015,
Kadlecova2017,Zitko2018,Kadlecova2019} and references therein.

For what follows, it is convenient to rewrite the previous Hamiltonian in terms of
four-dimensional spinors that live in a space resulting from the direct product
of the spin space and the Nambu (electron-hole) space. In the case of the leads,
the relevant spinor is defined as
\begin{equation}
\label{eq-c-tilde}
\tilde c^{\dagger}_{{\boldsymbol k}j} = \left( c^{\dagger}_{{\boldsymbol k}j\uparrow},
c_{{-\boldsymbol k}j\downarrow}, c^{\dagger}_{{\boldsymbol k}j\downarrow} ,
-c_{{-\boldsymbol k}j\uparrow} \right) ,
\end{equation}
while for the impurity states we define
\begin{equation}
\label{eq-d-tilde}
\tilde d^{\dagger} = \left( d^{\dagger}_{\uparrow}, d_{\downarrow}, 
d^{\dagger}_{\downarrow}, -d_{\uparrow} \right) .
\end{equation}

Using the notation $\tau_i$ and $\sigma_i$ ($i=1,2,3$) for Pauli matrices in Nambu and
spin space, respectively, and with $\tau_0$ and $\sigma_0$ as the unit matrices in those
spaces, it is straightforward to show that the Hamiltonian in Eq.~(\ref{eq-total-H}) can 
be cast into the form
\begin{subequations}
\begin{eqnarray}
H_j & = & \frac{1}{2} \sum_{\boldsymbol k} \tilde c^{\dagger}_{{\boldsymbol k}j} 
\hat H_{{\boldsymbol k} j} \tilde c_{{\boldsymbol k}j} , \\
H_{\rm i} & = & \frac{1}{2} \tilde d^{\dagger} \hat H_{\rm i} \tilde d ,	 \\
H_{\rm hopping} & = & \frac{1}{2} \sum_{{\boldsymbol k},j} \left\{
\tilde c^{\dagger}_{{\boldsymbol k}j} \hat V_{j,{\rm i}} \tilde d +
\tilde d^{\dagger} \hat V_{{\rm i},j} \tilde c_{{\boldsymbol k}j} \right\} ,
\end{eqnarray}
\end{subequations}
where
\begin{subequations}
\begin{eqnarray}
\hat H_{{\boldsymbol k} j} & = & \sigma_0 \otimes (\xi_{\boldsymbol k} \tau_3 + 
\Delta_j e^{i \varphi_j \tau_3} \tau_1 ), \\
\hat H_{\rm i} & = & U (\sigma_0 \otimes \tau_3) + J (\sigma_3 \otimes \tau_0) , \\
\hat V_{j,{\rm i}} & = & t_j(\sigma_0 \otimes \tau_3) = 
\hat V^{\dagger}_{{\rm i},j} .
\end{eqnarray}
\end{subequations}

\subsection{Bare Green's functions}

The starting point for the calculation of the electronic transport properties will
be the bare Green's functions of the different subsystems which can be easily calculated 
from the previous matrix Hamiltonians as follows. First, the retarded and advanced Green's
functions of the leads resolved in $\boldsymbol k$-space are defined as 
$\hat g^\mathrm{r,a}_{{\boldsymbol k},jj}(E) = (E \pm i \eta - \hat H_{{\boldsymbol k} j})^{-1}$, 
where $j= {\rm t, S}$ and $\eta=0^+$ is a positive infinitesimal parameter, which we 
shall drop out to simplify things along with the superscript $\mathrm{r,a}$, unless they are 
strictly necessary. Summing over ${\boldsymbol k}$, $\hat g_{jj} = \sum_{\boldsymbol k} 
\hat g_{{\boldsymbol k},jj} \rightarrow N_{0,j} \int^{\infty}_{-\infty} d 
\xi_{\boldsymbol k} \hat g_{jj}(\xi_{\boldsymbol k})$, where $N_{0,j}$ is the normal
density of states at the Fermi energy of lead $j$, we arrive at the standard expression
for the bulk Green's function of a BCS superconductor
\begin{equation}
\label{eq-gBCS}
\hat g_{jj}(E) = \frac{-\pi N_{0,j}}{\sqrt{\Delta^2_j - E^2}} \sigma_0 \otimes\left[E \tau_0 + 
\Delta_j e^{i \varphi_j \tau_3} \tau_1 \right] .  
\end{equation}

On the other hand, the impurity Green's function is given by
\begin{eqnarray}
\label{eq-gii}
\hat g_{\rm i i}(E) & = & (E- \hat H_{\rm i})^{-1} \nonumber \\
& = & \left( \begin{array}{cccc}
\frac{1}{E-U-J} & 0 & 0 & 0 \\ 0 & \frac{1}{E+U-J} & 0 & 0 \\ 
0 & 0 & \frac{1}{E-U+J} & 0 \\ 0 & 0 & 0 & \frac{1}{E+U+J} \end{array} 
\right).	
\end{eqnarray}

Given that in an experimental setup the coupling of the impurity to the STM tip is
weaker than the coupling to the substrate, a natural division of the system, which
will be very useful in the transport calculations, is to consider the tip on one
side and the impurity-substrate system on the other side. The dressed Green's 
function of the impurity taking into account the coupling to the superconducting 
substrate is calculated by solving the Dyson equation
\begin{equation}
\hat G_{jk} = \hat g_{jk} + \sum_{\alpha \beta} \hat g_{j \alpha}
\hat V_{\alpha \beta} \hat G_{\beta k} ,	
\end{equation}
where the indices run over ${\rm i}$ and ${\rm S}$, the bare Green's functions 
$\hat g_{\rm ii}$ and $\hat g_{\rm SS}$ are given by Eqs.~(\ref{eq-gii}) and (\ref{eq-gBCS}),
respectively, and the couplings are given by $\hat V_{\rm iS} = t_{\rm S} (\sigma_0
\otimes \tau_3) = \hat V^{\dagger}_{\rm Si}$. This Dyson equation can be 
easily solved to obtain
\begin{equation}
\label{eq-Gii}
\hat G_{\rm ii}(E) = \left( \begin{array}{cc} \hat G_{{\rm ii},\uparrow \uparrow}(E) & 0 \\
0 & \hat G_{{\rm ii},\downarrow \downarrow}(E) \end{array} \right) ,
\end{equation}
with
\begin{widetext}
\begin{equation}
\hat G_{{\rm ii},\sigma \sigma}(E) = \frac{1}{D_{\sigma}(E)} \left( \begin{array}{cc}
 E \Gamma_{\rm S} + (E+U-J_{\sigma}) \sqrt{\Delta^2_{\rm S} - E^2} & 
 \Gamma_{\rm S} \Delta_{\rm S} e^{i \varphi_{\rm S}} \\
 \Gamma_{\rm S} \Delta_{\rm S} e^{- i \varphi_{\rm S}} &
 E\Gamma_{\rm S} + (E-U-J_{\sigma}) \sqrt{\Delta^2_{\rm S} - E^2} \end{array} \right) ,
\end{equation}
\end{widetext}
where 
\begin{eqnarray}
\label{eq-Dup}
D_{\sigma}(E) & = & 2\Gamma_{\rm S} E (E-J_{\sigma}) +  \nonumber \\
& & \left[(E-J_{\sigma})^2 - U^2 -
\Gamma^2_{\rm S} \right] \sqrt{\Delta^2_{\rm S} - E^2} .
\end{eqnarray}
Here, $J_{\uparrow} = + J$ and $J_{\downarrow} = -J$ and we have defined the tunneling rate 
$\Gamma_{\rm S} = \pi N_{0, \rm S} t^2_{\rm S}$ (a similar rate $\Gamma_{\rm t} = 
\pi N_{0, \rm t} t^2_{\rm t}$ describes the strength of the tip-impurity coupling).

\subsection{Computing the current-voltage characteristics}

To compute the electronic transport properties in our model system, we shall assume that
the voltage drops at the interface between the impurity and the STM tip, which is justified
by the fact that usually the tip-impurity coupling is clearly weaker than the substrate-impurity
coupling. As mentioned above, with this assumption the natural division of the system to apply 
the standard nonequilibrium techniques is such that the left subsystem is the STM tip and the
right system is the combination of the magnetic impurity and the superconducting substrate. 
With this division, we can treat our system as a single-channel point contact in which the 
effective Hamiltonian reads \cite{Cuevas1996}
\begin{equation}
\label{eq-HLR}
H = H_\mathrm{L} + H_\mathrm{R} + \sum_{\sigma} t_{\rm t} \left\{ e^{i \varphi(t)/2} 
c^{\dagger}_{\rm L\sigma} c_{\rm R\sigma} + e^{-i\varphi(t)/2} 
c^{\dagger}_{\rm R \sigma} c_{\rm L \sigma} \right\},	
\end{equation}
where $H_\mathrm{L}$ is now the BCS Hamiltonian of the tip, $H_\mathrm{R}$ is the Hamiltonian 
of the impurity coupled to the substrate, $t_{\rm t}$ is the tunneling rate describing the
coupling between the impurity and the tip, and $\varphi(t) = \varphi_0 + 2eVt/\hbar$ is the 
time-dependent superconducting phase difference, with $V$ being the applied voltage and 
$\varphi_0 = \varphi_\mathrm{L} - \varphi_\mathrm{R}$ is the dc part of this phase difference. 
In the coupling term of this Hamiltonian L and R stand for the outermost sites of each electrode
and, in particular, R corresponds now to the dressed impurity site. With this starting point, the
calculation of the current-voltage characteristics can be done following the theory of MARs described
in Ref.~\cite{Cuevas1996}, where things need to be slightly modified to accommodate for the
breaking of the spin degeneracy in this case, as we describe in what follows. 

First, we evaluate the current at the interface between the two electrodes (L and R),
which adopts the form
\begin{eqnarray}
I(t) & = & \frac{ie}{\hbar} \sum_{\sigma} t_{\rm t} \left\{ e^{i \varphi(t)/2} \langle 
c^{\dagger}_{\rm L\sigma}(t) c_{\rm R\sigma}(t) \rangle - \right. \nonumber \\ 
& & \hspace{1.5cm} \left.
e^{-i\varphi(t)/2} \langle c^{\dagger}_{\rm R \sigma}(t) c_{\rm L \sigma}(t) \rangle \right\}.
\end{eqnarray}
The nonequilibrium expectation values appearing in the previous equation can be written
in terms of the lesser Green's functions $\hat G^{+-}_{jk}$ which in the $4 \times 4$
spin-Nambu representation are given by
\begin{equation}
\hat G^{+-}_{jk}(t,t^{\prime}) = -i \langle T_\mathrm{C} \left\{ 
 \tilde c_j(t_{+}) \tilde c^{\dagger}_k(t^{\prime}_{-}) \right\} \rangle .	
\end{equation}
Here, $T_\mathrm{C}$ is the time-ordering operator on the Keldysh contour such that any
time in the lower branch ($t^{\prime}_{-}$) is larger than any time in the upper one 
($t_{+}$). Moreover, $j,k=\mathrm{L,R}$ and the four-component spinors $\tilde c_j$ and
$\tilde c^{\dagger}_k$ are defined following the conventions of Eqs.~(\ref{eq-c-tilde}) 
and (\ref{eq-d-tilde}). Thus, the current can now be written as
\begin{eqnarray}
I(t) & = & \frac{e}{2\hbar} \mbox{Tr} \left\{ (\sigma_0 \otimes \tau_3) \left[ 
\hat v_\mathrm{LR}(t) \hat G^{+-}_\mathrm{RL}(t,t) - \right. \right. \nonumber \\ 
& & \left. \left. \hspace{2.6cm} \hat v_\mathrm{RL}(t) 
\hat G^{+-}_\mathrm{LR}(t,t) \right] \right\}, 
\end{eqnarray}
where $\mbox{Tr}$ is the trace taken over Nambu and spin degrees of freedom and  
$\hat v_\mathrm{LR}(t) = t_{\rm t} (\sigma_0 \otimes \tau_3 e^{i\varphi(t) \tau_3/2} ) = 
\hat v^{\dagger}_\mathrm{RL}(t)$ are the coupling matrices in spin-Nambu space.

To determine the dressed Green's functions appearing in the current formula
we follow a perturbative scheme and treat the coupling term in the Hamiltonian of 
Eq.~(\ref{eq-HLR}) as a perturbation. The unperturbed Green's functions, $\hat g$, 
correspond to the uncoupled electrodes in equilibrium. To be precise, the bare 
Green's function $\hat g_\mathrm{LL}$ is given by Eq.~(\ref{eq-gBCS}) and the bare 
Green's function $\hat g_\mathrm{RR}$ of the impurity coupled to the substrate is given 
by Eq.~(\ref{eq-Gii}) [without the superconducting phase that has been moved to the
couplings]. On the other hand, it is convenient to express the current by means of 
the so-called T-matrix. The T-matrix associated to the time-dependent perturbation 
in Eq.~(\ref{eq-HLR}) is defined as
\begin{equation}
\hat T^\mathrm{r,a} = \hat v + \hat v \circ \hat g^\mathrm{r,a} \circ \hat T^\mathrm{r,a} ,
\end{equation}
where the $\circ$ product is a shorthand for convolution, i.e., for integration over 
intermediate time arguments. As shown in Ref.~\cite{Cuevas1996}, the exact current 
including all the orders in the tunneling rate can be written in terms of the T-matrix 
components as
\begin{widetext}
\begin{eqnarray}
I(t) & = & \frac{e}{2\hbar} 	\mbox{Tr} \left\{ (\sigma_0 \otimes \tau_3) \left[
\hat T^\mathrm{r}_\mathrm{LR} \circ \hat g^{+-}_\mathrm{RR} \circ 
\hat T^\mathrm{a}_\mathrm{RL} \circ \hat g^\mathrm{a}_\mathrm{LL}
- \hat g^\mathrm{r}_\mathrm{LL} \circ \hat T^\mathrm{r}_\mathrm{LR} \circ 
\hat g^{+-}_\mathrm{RR} \circ \hat T^\mathrm{a}_\mathrm{RL} + 
\right. \right. \nonumber \\ & & \left. \left. \hspace{2.7cm}  
\hat g^\mathrm{r}_\mathrm{RR} \circ \hat T^\mathrm{r}_\mathrm{RL} \circ 
\hat g^{+-}_\mathrm{LL} \circ \hat T^\mathrm{a}_\mathrm{LR} - 
\hat T^\mathrm{r}_\mathrm{RL} \circ \hat g^{+-}_\mathrm{LL} \circ 
\hat T^\mathrm{a}_\mathrm{LR} \circ \hat g^\mathrm{a}_\mathrm{RR}
\right] \right\} .
\end{eqnarray}
\end{widetext}

In order to solve the T-matrix integral equations it is convenient to Fourier
transform with respect to the temporal arguments
\begin{equation}
\hat T(t,t^{\prime}) = \frac{1}{2\pi} \int^{\infty}_{-\infty} dE 
\int^{\infty}_{-\infty} dE^{\prime} e^{-iEt} e ^{iE^{\prime} t^{\prime}}
\hat T(E,E^{\prime}) .	
\end{equation}
Because of the time dependence of the coupling matrices, one can show that
$\hat T(E,E^{\prime})$ admits the following general solution
\begin{equation}
\hat T(E,E^{\prime}) = \sum_n \hat T(E, E +neV) \delta(E-E^{\prime} + neV) .
\end{equation}
Thus, one can show that the current has the time dependence
\begin{equation}
\label{eq-It}
I(t) = \sum_n I_n e^{i n \varphi(t)} ,	
\end{equation}
where the current amplitudes $I_n$ can be expressed in terms of the T-matrix 
Fourier components, $\hat T_{nm}(E) \equiv \hat T(E+neV,E+meV)$, as
\begin{widetext}
\begin{eqnarray}
I_n & = & \frac{e}{2h} \int^{\infty}_{-\infty} dE \sum_m 
\mbox{Tr} \left\{ (\sigma_0 \otimes \tau_3) \left[	
\hat T^\mathrm{r}_{\mathrm{LR},0m} \hat g^{+-}_{\mathrm{RR},m} 
\hat T^\mathrm{a}_{\mathrm{RL},mn} \hat g^\mathrm{a}_{\mathrm{LL},n}
- \hat g^\mathrm{r}_{\mathrm{LL},0} \hat T^\mathrm{r}_{\mathrm{LR},0m} 
\hat g^{+-}_{\mathrm{RR},m} \hat T^\mathrm{a}_{\mathrm{RL},mn} +
\right. \right. \nonumber \\ & & \left. \left. \hspace{4.6cm}
\hat g^\mathrm{r}_{\mathrm{RR},0} \hat T^\mathrm{r}_{\mathrm{RL},0m} 
\hat g^{+-}_{\mathrm{LL},m} \hat T^\mathrm{a}_{\mathrm{LR},mn} - 
\hat T^\mathrm{r}_{\mathrm{RL},0m} \hat g^{+-}_{\mathrm{LL},m} 
\hat T^\mathrm{a}_{\mathrm{LR},mn} \hat g^\mathrm{a}_{\mathrm{RR},n}
\right] \right\} ,
\end{eqnarray}
\end{widetext}
where we have used the notation $\hat g_{jj,n}(E) = 
\hat g_{jj}(E+neV)$, notice that the bare Green's functions are diagonal in 
energy space, and the bare lesser Green's functions are given by $\hat g^{+-}_{jj}(E) =
\left[ \hat g^\mathrm{a}_{jj}(E) - \hat g^\mathrm{r}_{jj}(E) \right] f(E)$, where 
$f(E) = \left[ 1 + \exp(E/k_\mathrm{B}T) \right]^{-1}$ is the Fermi function with 
$T$ being the temperature. The previous formula can be further simplified by using 
the general relation $\hat T^\mathrm{r,a}_{\mathrm{RL},nm}(E) = 
(\hat T^\mathrm{a,r}_{\mathrm{LR},mn})^{\dagger}(E)$, which reduces the calculation
of the current to the determination of the Fourier components 
$\hat T^\mathrm{r,a}_{\mathrm{LR},nm}$ fulfilling the following set of linear 
algebraic equations
\begin{eqnarray}
\label{eq-T}
\hat T^\mathrm{r,a}_{\mathrm{LR},nm} & = & \hat v_{\mathrm{LR},nm} + 
\hat {\cal E}^\mathrm{r,a}_n \hat T^\mathrm{r,a}_{\mathrm{LR},nm} +
\nonumber \\ & & \hat {\cal V}^\mathrm{r,a}_{n,n-2} \hat T^\mathrm{r,a}_{\mathrm{LR},n-2,m} + 
\hat {\cal V}^\mathrm{r,a}_{n,n+2} \hat T^\mathrm{r,a}_{\mathrm{LR},n+2,m} ,
\end{eqnarray}
where the different matrix coefficients are given in terms of the unperturbed Green's
functions as follows
\begin{eqnarray} 
\hat v_{\mathrm{LR},nm} & = & \frac{t_{\rm t}}{2} \sigma_0 \otimes \left[ 
(1 + \tau_3) \delta_{m,n+1} - (1 - \tau_3) \delta_{m,n-1} \right] , \nonumber \\
\hat v_{\mathrm{RL},nm} & = & \frac{t_{\rm t}}{2} \sigma_0 \otimes \left[ 
(1 + \tau_3) \delta_{m,n-1} - (1 - \tau_3) \delta_{m,n+1} \right] , \nonumber \\
\hat {\cal E}^\mathrm{r,a}_n & = & \hat v_{LR,n,n+1} \hat g^\mathrm{r,a}_{\mathrm{RR},n+1} 
\hat v_{\mathrm{RL},n+1,n} \hat g^\mathrm{r,a}_{\mathrm{LL},n} + \nonumber \\
& & \hat v_{\mathrm{LR},n,n-1} \hat g^\mathrm{r,a}_{\mathrm{RR},n-1} 
\hat v_{\mathrm{RL},n-1,n} \hat g^\mathrm{r,a}_{\mathrm{LL},n} ,
\nonumber \\
\hat {\cal V}^\mathrm{r,a}_{n,n-2} & = & \hat v_{\mathrm{LR},n,n-1} 
\hat g^\mathrm{r,a}_{\mathrm{RR},n-1} \hat v_{\mathrm{RL},n-1,n-2} 
\hat g^\mathrm{r,a}_{\mathrm{LL},n-2} , \nonumber \\
\hat {\cal V}^\mathrm{r,a}_{n,n+2} & = & \hat v_{\mathrm{LR},n,n+1} 
\hat g^\mathrm{r,a}_{\mathrm{RR},n+1} \hat v_{\mathrm{RL},n+1,n+2} 
\hat g^\mathrm{r,a}_{\mathrm{LL},n+2} .
\end{eqnarray}
In general, these block-tridiagonal systems have to be solved numerically and 
the current can only be expressed in an analytical form in a few limiting 
cases, as we discuss below. On the other hand, let us stress that we shall focus
here exclusively on the discussion of the dc current, i.e., $I_0$ in 
Eq.~(\ref{eq-It}), and we shall not analyze the (zero-bias) dc Josephson current. 
To conclude this discussion let us say that the $4 \times 4$ formalism presented
here is not strictly necessary in the case of a single impurity with spin-preserving
tunneling, and the transport properties of our model system can be equivalently
described within a $2 \times 2$ formalism (using only the Nambu space). However, 
the $4 \times 4$ approach provides a convenient platform that illustrates the key
role of the spin and it becomes absolutely necessary in more complex situations
like, for instance, those involving the tunneling between magnetic impurities
with noncollinear spins \cite{Huang2019b}.

\subsection{YSR states and tunneling spectra} \label{sec-tunneling-spectra}

Obviously, the transport properties will reflect the electronic structure of the 
magnetic impurity. In particular, due to the coupling of the impurity with the
superconducting substrate, this electronic structure will exhibit both Andreev 
and YSR bound states in the gap region. In the limit in which we can ignore the 
coupling to the STM tip ($\Gamma_{\rm t} = 0$), the local density of states (LDOS) 
projected onto the impurity site is given by 
\begin{equation}
\label{eq-LDOS-imp}	
\rho_{\rm Total,imp}(E) = \rho_{\uparrow}(E) + \rho_{\downarrow}(E), 
\end{equation}
where
\begin{equation}
\label{eq-LDOS-imp-sigma}
\rho_{\sigma}(E) = \frac{1}{\pi} \mbox{Im} \left\{ \frac{E \Gamma_{\rm S} + 
(E+U-J_{\sigma}) \sqrt{\Delta^2_{\rm S} - E^2}}{D_{\sigma}(E)} \right\} , 	
\end{equation}
where $E=E-i\eta_{\rm S}$ and $D_{\sigma}(E)$ is given by Eq.~(\ref{eq-Dup}). 
Here, $\eta_{\rm S}$ is a phenomenological parameter that describes the inelastic 
broadening of the electronic states. The condition for the appearance of superconducting
bound states is $D_{\sigma}(E) = 0$. In particular, the spin-induced YSR states appear 
in the limit $J \gg |\Delta_\mathrm{S}|$ (and they are inside the gap when also
$\Gamma_\mathrm{S} \gg \Delta_\mathrm{S}$). In this case, there is a pair of fully 
spin-polarized YSR bound states at energies (measured with respect to the Fermi energy)
\cite{Huang2019a}
\begin{equation}
\label{eq-YSR1}
\epsilon_{\rm S} = \pm \Delta_{\rm S} \frac{J^2 - \Gamma^2_{\rm S} - U^2}
{\sqrt{ \left[ \Gamma^2_{\rm S} + (J-U)^2 \right] 
\left[ \Gamma^2_{\rm S} + (J+U)^2 \right]}} ,
\end{equation}
which has a similar structure as in the case of the classical Shiba model 
\cite{Salkola1997,Flatte1997a,Flatte1997b,Balatsky2006}. This is more apparent 
in the electron-hole symmetric case $U=0$, where the previous expression reduces to 
\begin{equation}
\label{eq-YSR2}
\epsilon_{\rm S} = \pm \Delta_{\rm S} \frac{J^2 - \Gamma^2_{\rm S}}
{J^2 + \Gamma^2_{\rm S}} . 
\end{equation}

\begin{figure}[t]
\includegraphics[width=\columnwidth,clip]{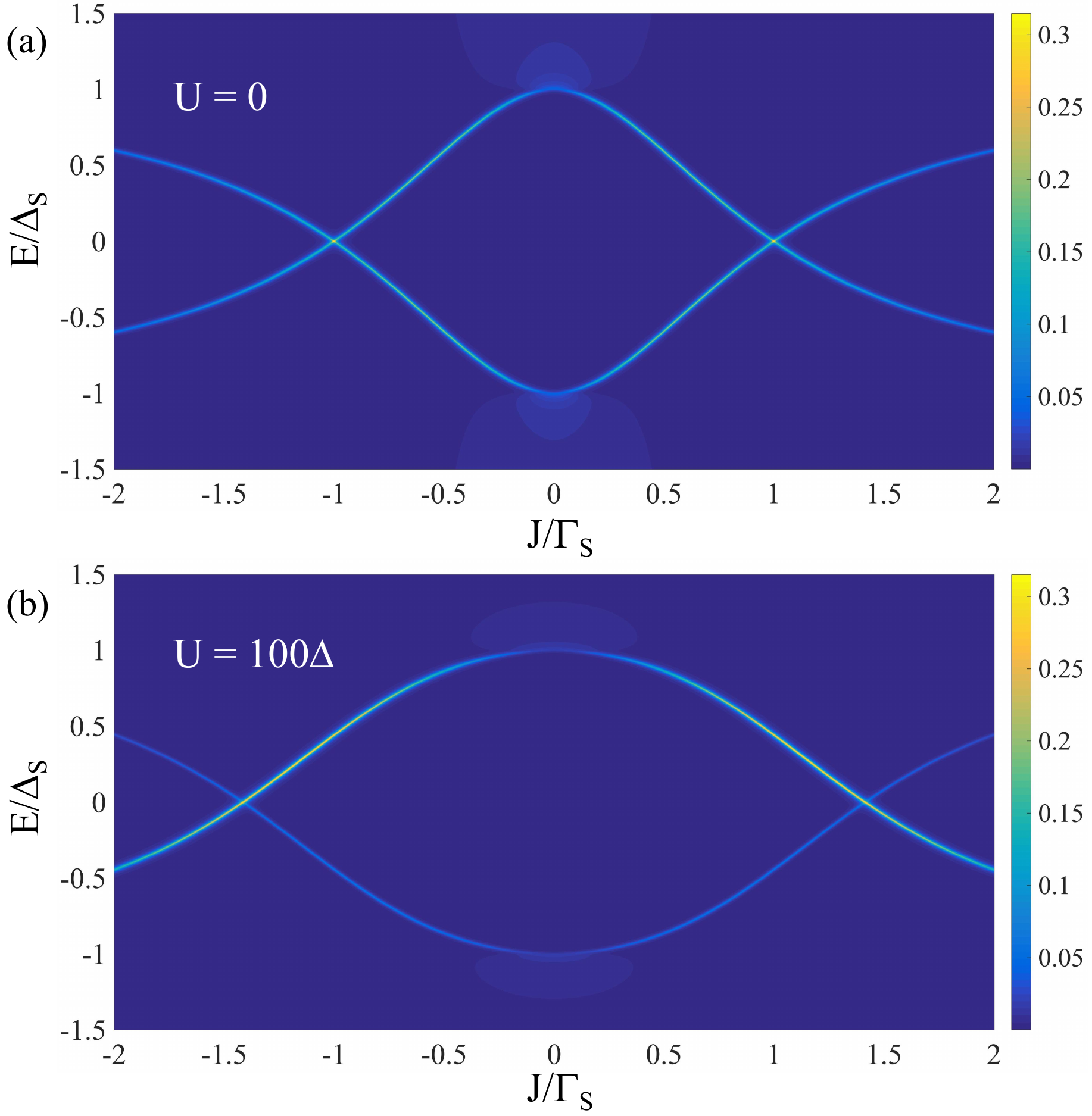}
\caption{(a) Local density of states (LDOS) projected onto the magnetic impurity 
in units of $1/\Delta$ as a function of the energy $E$ and the exchange energy $J$ 
for a situation in which the impurity is uncoupled to the STM tip. The different 
parameters are: $\Gamma_{\rm t} = 0$, $\Gamma_{\rm S} = 100\Delta_{\rm S}$, $U = 0$, 
and $\eta_{\rm S} = 0.01\Delta_{\rm S}$. (b) The same as in panel (a) but for 
$U = 100\Delta_{\rm S}$.}
\label{fig-LDOS}
\end{figure}

For future reference, we present in Fig.~\ref{fig-LDOS} two representative 
cases of the LDOS in the impurity when it is only coupled to the substrate
for a case with electron-hole symmetry (upper panel) and a case in which this
symmetry is broken (lower panel). Notice the appearance of a pair of YSR 
states inside the gap whose dependence on the exchange energy is accurately
described by the analytical formula of Eq.~(\ref{eq-YSR1}). Notice also that
the two states cross at zero energy when $J^2 = \Gamma^2_{\rm S} + U^2$, which
is the point that corresponds to the quantum critical point \cite{Balatsky2006}. 
Finally, it is also worth noticing the absence of singularities at 
$E=\pm \Delta_{\rm S}$, which is a simple consequence of the appearance of the 
YSR states and the conservation of the number of states.

\begin{figure}[t]
\includegraphics[width=\columnwidth,clip]{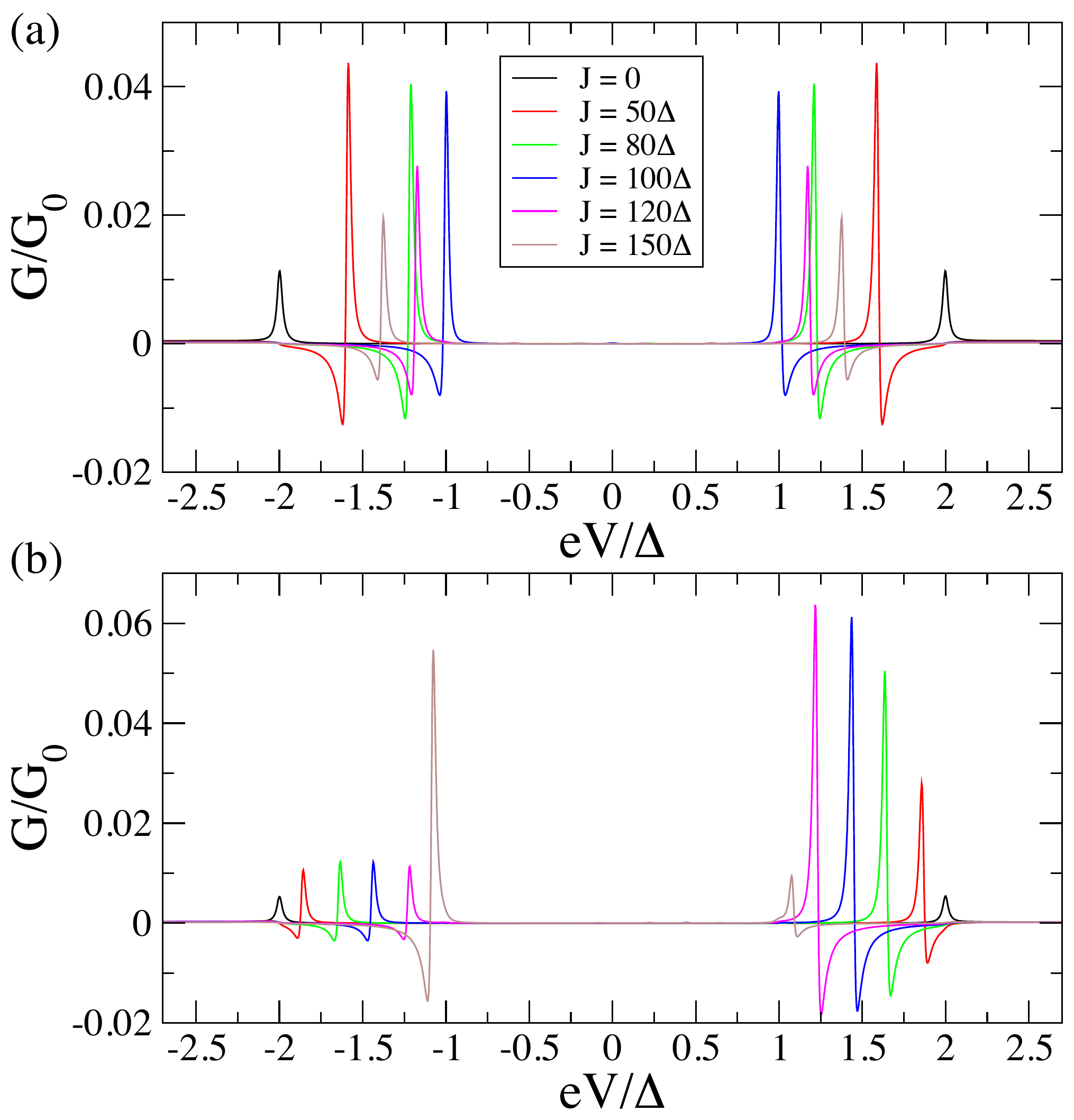}
\caption{(a) Differential conductance $G=dI/dV$ as a function of the bias voltage in the 
tunnel regime for a superconducting STM tip and for different values of the exchange 
energy, as indicated in the legend. The energy gaps of the tip and the substrate are
assumed to be equal: $\Delta_{\rm t} = \Delta_{\rm S} = \Delta$. The rest of the parameters 
have the values: $\Gamma_{\rm t} = 0.01\Delta$, $\Gamma_{\rm S} = 100\Delta$, $U = 0$, 
$\eta_{\rm t} = \eta_{\rm S} = 0.01\Delta$, and $k_{\rm B}T = 0.01\Delta$. 
(b) The same as in panel (a), but for $U = 100\Delta$.}
\label{fig-tunneling-spectra}
\end{figure}

As a next step, we briefly remind how the presence of the YSR states is reflected
in the tunneling spectra acquired with a superconducting tip, i.e., when the
tip is sufficiently far away such that $\Gamma_{\rm t} \ll \Gamma_{\rm S}$ and the
only relevant tunneling process is the single-quasiparticle tunneling. In this case, 
one use the approximation $\hat T^\mathrm{r,a}_{\mathrm{LR},nm} \approx \hat 
v_{\mathrm{LR},nm}$ in Eq.~(\ref{eq-T}) to arrive at the following analytical expression
\begin{multline}
I_{\rm tunnel}(V) = \frac{4 \pi^2 \Gamma_{\rm t} e}{h} 
\int^{\infty}_{-\infty} dE \, \tilde \rho_{\rm tip}(E-eV)  \\ 
\times \rho_{\rm Total,imp}(E)  \left[ f(E-eV) - f(E) \right] , 
\end{multline}
where $f(E)$ is the Fermi function, $\tilde \rho_{\rm tip}$ is the dimensionless BCS DOS 
of the tip, and $\rho_{\rm Total,imp}(E)$ is total (including both spin contributions) LDOS
in the impurity given by Eqs.~(\ref{eq-LDOS-imp}) and (\ref{eq-LDOS-imp-sigma}). In 
Fig.~\ref{fig-tunneling-spectra} we illustrate the lineshapes of the differential
conductance $dI/dV$ (in units of the quantum of conductance $G_0=2e^2/h$) in the
tunnel regime for different values of the exchange energy $J$ in the impurity and 
for two different values of impurity on-site energy: $U = 0$ and $U = 100 \Delta$.
Here, we assume that the gaps of the tip and substrate are equal: $\Delta_{\rm t} = 
\Delta_{\rm S} = \Delta$. Notice that in all cases the most prominent feature is the
appearance of a conductance peak at $eV = \pm (\Delta + |\epsilon_{\rm S}|)$, which is 
accompanied by a negative differential conductance (except in the case $J=0$ where
there are no YSR states and one recovers the standard coherent peaks at $eV = \pm 2\Delta$). 
Notice also that the differential conductance is symmetric, i.e., independent of the bias
polarity, when there is electron-hole symmetry, see panel (a), and asymmetric when the
electron-hole symmetry is broken, see panel (b). It is also worth remarking the absence of
coherent peaks at $eV= \pm 2\Delta$ when there is no spin degeneracy, i.e., $J \neq 0$, 
which is due to the absence of singularities at the gap edges in the impurity LDOS. 

\subsection{Normal state conductance}

In what follows, and in order to make contact with the experiment, it is important to
characterize our system with the normal state conductance, $G_{\rm N}$. In the case
in which neither the tip nor the substrate are superconducting, the current formula 
within our model can be worked out analytically and it is given by the following 
Landauer-type of expression
\begin{multline}
I_{\rm normal}(V) = \frac{e}{h} \int^{\infty}_{-\infty} dE \, 
\left\{ \tau_{\uparrow}(E) + \tau_{\downarrow}(E) \right\}  \\ 
\times \left[ f(E-eV) - f(E) \right] , 
\end{multline}
where the spin-dependent transmission coefficients are given by
\begin{equation}
\tau_{\sigma}(E) = \frac{4\Gamma_{\rm t} \Gamma_{\rm S}}
{(E-U-J_{\sigma})^2 + (\Gamma_{\rm t} + \Gamma_{\rm S})^2} .
\end{equation}
Thus, the zero-temperature normal state linear conductance is given by
\begin{equation}
\label{eq-GN}
\frac{G_{\rm N}}{G_0} =  \frac{1}{2} \left\{ \tau_{\uparrow}(E=0) +
\tau_{\downarrow}(E=0) \right\} ,
\end{equation}
which in the tunnel regime ($\Gamma_\mathrm{t} \ll \Gamma_\mathrm{S}$) reduces to
\begin{equation}
\frac{G_{\rm N}}{G_0} \approx  
\frac{2\Gamma_{\rm t} \Gamma_{\rm S}} {(U+J)^2 + \Gamma^2_{\rm S}} +
\frac{2\Gamma_{\rm t} \Gamma_{\rm S}} {(U-J)^2 + \Gamma^2_{\rm S}} ,
\end{equation}
showing that it is linear in $\Gamma_\mathrm{t}$.
Moreover, in this work, $|eV|$ will always be much smaller than
$\Gamma_{\rm t} + \Gamma_{\rm S}$ such that the differential conductance in 
the normal state will be independent of the bias.

\section{A normal tip: Andreev reflection mediated by YSR states} \label{sec-NS}

Since our central goal is the study of the interplay between YSR states
and Andreev reflections, it is convenient to first discuss the results 
for the current-voltage characteristics in the case in which the STM tip is
not superconducting. In this case, the total current is the sum of two 
contributions: the current due to single-quasiparticle tunneling and 
the current due to an Andreev reflection, which can be mediated by the
YSR states, as we shall show below. Assuming that the STM tip is in the normal 
state, the expression of the current can be worked out analytically
for arbitrary range of parameters and it adopts the form
\begin{equation}
I_{\rm NS}(V) = I_{\rm qp}(V) + I_{\rm Andreev}(V) ,
\end{equation}
where $I_{\rm qp}(V)$ is the quasiparticle current and $I_{\rm Andreev}(V)$
the Andreev current. While the expression of the quasiparticle current is too 
cumbersome and we shall not present it here explicitly, the Andreev current 
adopts a compact and intuitive form given by
\begin{multline}
	I_{\rm Andreev}(V) =  \frac{2e}{h}  \int^{\infty}_{-\infty} dE \, 
	R_{\rm A}(E)  \\
	\times \left[ f(E-eV) - f(E+eV) \right]  ,
\end{multline}
where the Andreev reflection probability is given by
\begin{equation}
R_{\rm A}(E) = \frac{2 \Gamma^2_{\rm t} \Gamma^2_{\rm S} 
\Delta^2_{\rm S}} {|D_{\rm A, \uparrow}(E)|^2} + 
\frac{2 \Gamma^2_{\rm t} \Gamma^2_{\rm S} \Delta^2_{\rm S}} 
{|D_{\rm A, \downarrow}(E)|^2} ,
\end{equation}
with
\begin{multline}
D_{\rm A, \sigma} (E) = 2\Gamma_{\rm S} E(E-J_{\sigma}-i\Gamma_{\rm t}) + \bigl[ (E-J_{\sigma})^2  \\
  - U^2 - \Gamma^2_{\rm S} - \Gamma^2_{\rm t} - 
2i\Gamma_{\rm t} (E-J_{\sigma}) \bigr] \sqrt{\Delta^2_{\rm S} - E^2} . 
\end{multline}
Notice that in the limit of weak coupling to the STM tip, $D_{\rm A, \sigma} (E) 
\approx D_{\sigma}(E)$, see Eq.~(\ref{eq-Dup}), where $D_{\sigma}(E)$ are the denominators 
whose zeros give us the energies of the YSR states in the case in which the impurity is 
not coupled to the STM tip.

From the expression above for $I_{\rm Andreev}(V)$, one can deduce several important
facts. First, since the Andreev reflection probability is always electron-hole 
symmetric, i.e., $R_{\rm A}(E) = R_{\rm A}(-E)$, the corresponding contribution of
this process to the differential conductance does not depend on the bias polarity: 
$G_\mathrm{A}(V) = G_\mathrm{A}(-V)$ (notice that at low temperatures 
$G_\mathrm{A}(V) = 2G_0 R_\mathrm{A}(eV))$. This fact was first recognized in 
Ref.~\cite{Martin2014}. This is at variance with the single-quasiparticle 
contribution, which does depend on the bias polarity if the electron-hole 
symmetry is broken ($U \neq 0$). On the other hand, the Andreev reflection 
probability is resonantly enhanced at the energy of the YSR states and therefore,
it gives rise to differential conductance peaks at $eV = \pm |\epsilon_{\rm S}|$.
This is similar to the feature expected from the quasiparticle current, whose
corresponding differential conductance increases significantly when the chemical
potential of the tip is aligned with the YSR states, i.e., also when $eV = \pm 
|\epsilon_{\rm S}|$. The way to differentiate between the contributions of quasiparticle 
tunneling and Andreev reflection is by studying how the differential conductance scales 
with the normal state conductance and by examining the dependence on the bias polarity. 

\begin{figure}[t]
\includegraphics[width=\columnwidth,clip]{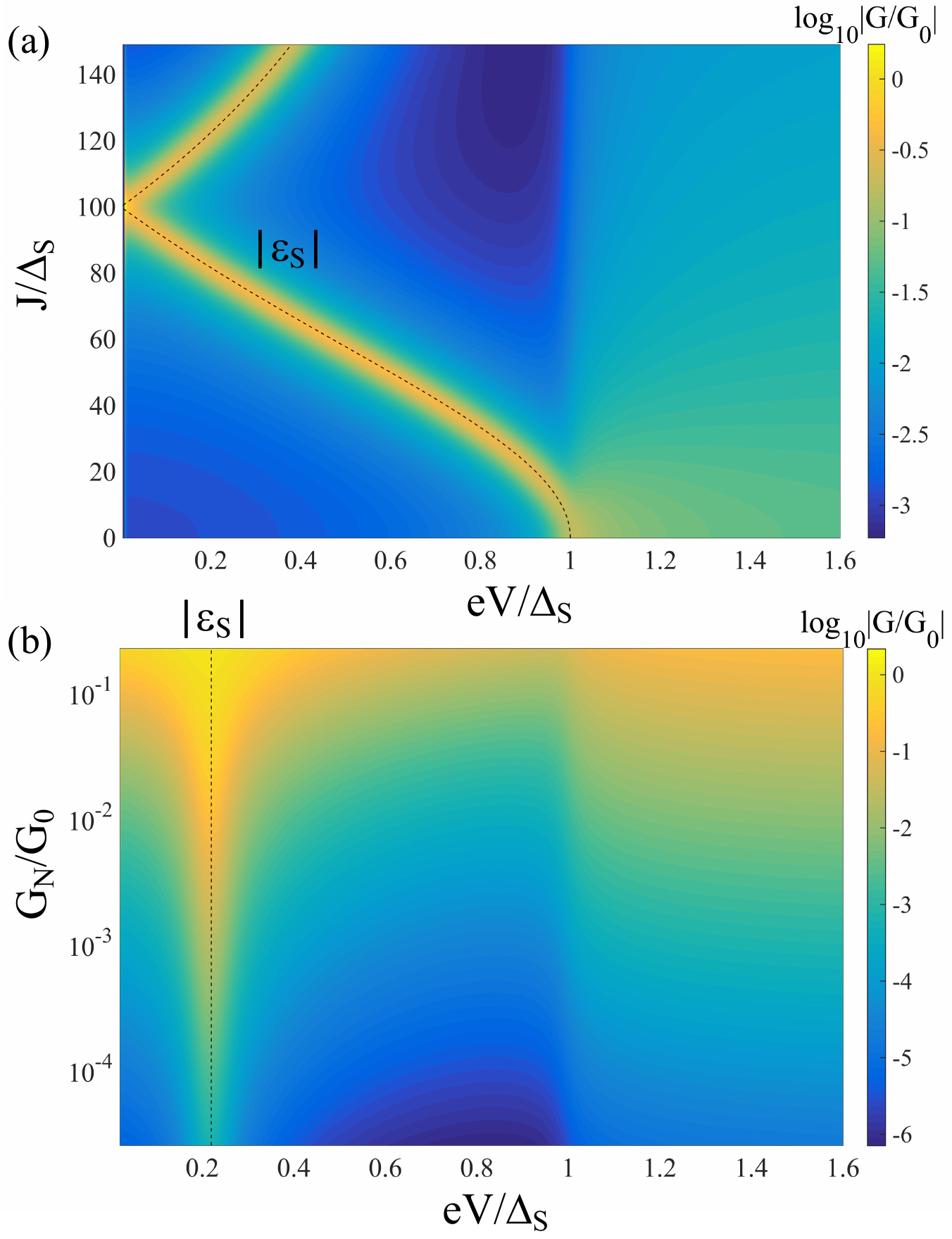}
\caption{(a) The logarithm (to the base 10) of the absolute value of the differential 
conductance $G=dI/dV$, normalized by the quantum of conductance $G_0=2e^2/h$, as a 
function of the bias voltage and the exchange energy for a non-superconducting STM tip 
($\Delta_{\rm t} = 0$). The values of the different parameters are: $\Gamma_{\rm S} = 
100\Delta_{\rm S}$, $\Gamma_{\rm t} = \Delta_{\rm S}$, $U = 0$, $\eta_{\rm S} = 
0.01\Delta_{\rm S}$, and $k_{\rm B}T = 0.01\Delta_{\rm S}$. The dashed line corresponds 
to the absolute value of the energy of the YSR bound state, as given by Eq~(\ref{eq-YSR2}). 
(b) The same as in panel (a), but as a function of the bias voltage and the normal state 
conductance $G_\mathrm{N}$ (in units of $G_0$) for $J = 80\Delta_{\rm S}$. The rest of 
the parameters have the same values as in panel (a). The vertical dashed line indicates 
the energy of YSR bound state, as computed from Eq.~(\ref{eq-YSR2}).}
\label{fig-NS1}
\end{figure}

We illustrate the expected results in the case of a normal conducting tip in
Fig.~\ref{fig-NS1}(a) where we show the differential conductance in a logarithmic scale 
as a function of the bias voltage and exchange energy for a case with a moderate
tunneling rate $\Gamma_{\rm t} = \Delta_{\rm S}$ and electron-hole symmetry ($U=0$), see 
figure caption for the value of the other model parameters. In this case the differential 
conductance is symmetric, $G(V)=G(-V)$, and for this reason we only show the region of 
positive bias. As expected, the most prominent feature is the appearance of a conductance
peak inside the gap ($eV \le \Delta_{\rm S}$) at a bias equal to the energy of the YSR 
state, which in this case is given by Eq~(\ref{eq-YSR2}).

While the dependence of the differential conductance on the exchange energy is very 
revealing, it is not easy to investigate experimentally in a continuous manner. 
In the experiments, it is much easier to control the normal state conductance, which
can be done by simply changing the tip-impurity distance. For this reason, we present
in Fig.~\ref{fig-NS1}(b) an example of the evolution of the differential conductance 
as a function of the normal state conductance, which we vary here by changing accordingly
the tunneling rate $\Gamma_{\rm t}$. In this case, we have chosen $J=80\Delta_{\rm S}$
and $U=0$, which makes the differential conductance independent of the bias polarity.
Again, the most salient feature is the appearance of a peak inside the gap at a bias
$eV = |\epsilon_{\rm S}| \approx 0.22\Delta_{\rm S}$. As the normal state conductance
increases, the peak first broadens and then at high transmissions ($G_{\rm N} \gtrsim
0.2G_0$) it becomes a very broad peak at zero bias (not shown here), in agreement 
with the experimental observations in Ref.~\cite{Brand2018} and the theoretical
discussion presented in that work.  

\begin{figure}[t]
\includegraphics[width=\columnwidth,clip]{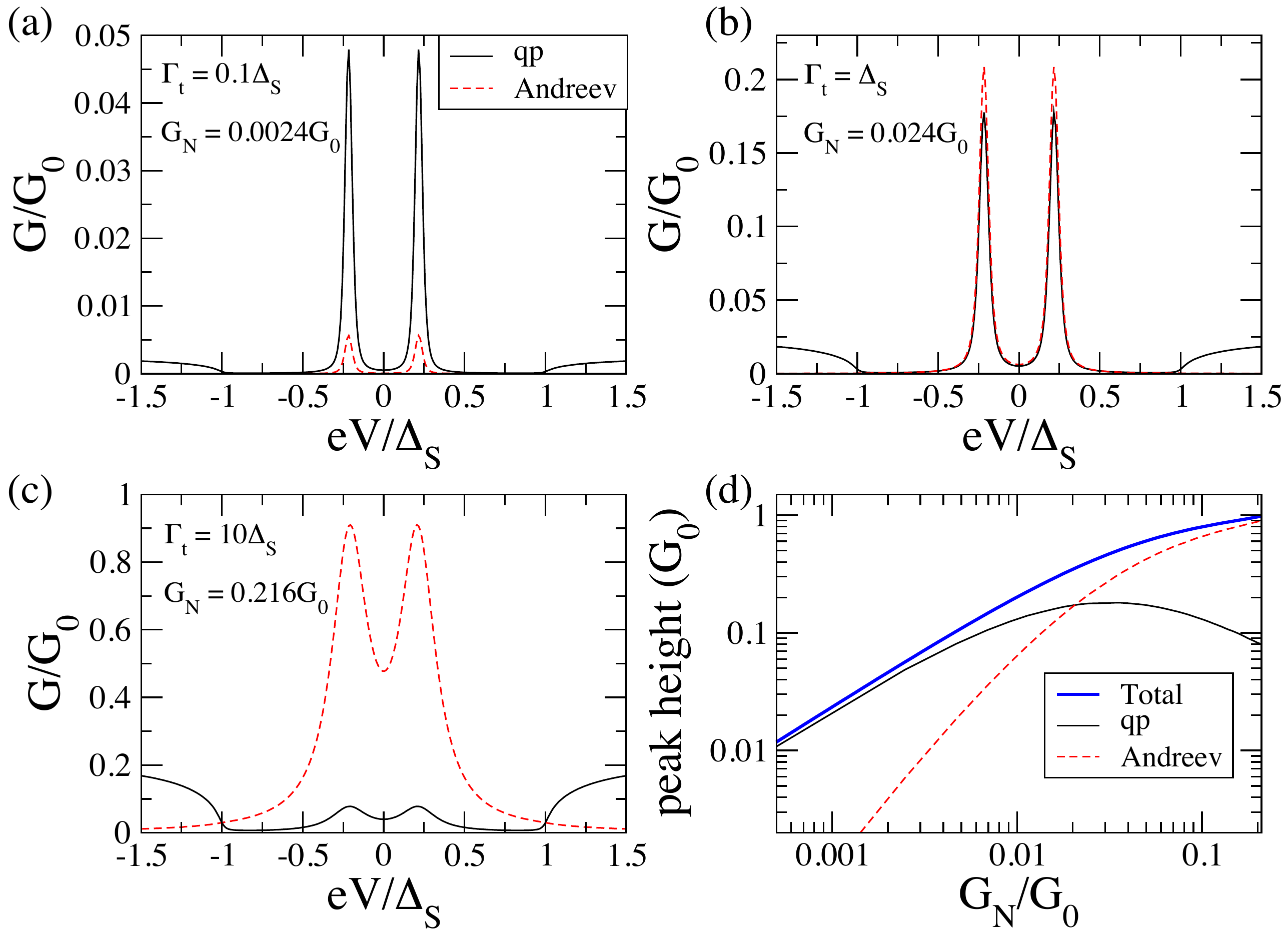}
\caption{Panels (a-c) show for a non-superconducting STM tip the individual contributions 
to the differential conductance of single-quasiparticle tunneling (black solid line) and 
Andreev reflection (red dashed line) for three different values of the tunneling rate 
$\Gamma_t$ (and normal state conductance), as indicated in the legend. The values of the 
different parameters are: $\Gamma_{\rm S} = 100\Delta_{\rm S}$, $J=80\Delta_{\rm S}$, 
$U = 0$, $\eta_{\rm S} = 0.01\Delta_{\rm S}$, and $k_{\rm B}T = 0.01\Delta_{\rm S}$.
(d) The value of the conductance peak, in units of $G_0$, as a function of the 
normal state conductance for the same parameters as in the other three panels.
The blue solid line corresponds to the total contribution, the black solid line
to the contribution from single-quasiparticle tunneling and the red dashed line
to the Andreev contribution.}
\label{fig-NS2}
\end{figure}

At this stage, the most relevant question concerns the relative contributions to the 
conductance peak inside the gap of quasiparticle tunneling and Andreev reflection. 
To answer this question we present in Fig.~\ref{fig-NS2}(a-c) these two individual 
contributions to the differential conductance for three representative values of 
$\Gamma_{\rm t}$ (or $G_{\rm N}$) corresponding to the results of Fig.~\ref{fig-NS1}(b). 
As one can see, when the normal state conductance (or normal transmission) is sufficiently
small, the subgap differential conductance is dominated by the contribution of 
single-quasiparticle tunneling, as expected. However, as the normal state transmission 
increases, the Andreev contribution becomes more relevant and eventually, it dominates the
subgap conductance and, in particular, the conductance peak at the energy of the YSR state.
In this particular
example both contributions become of the same order when $G_{\rm N} \approx 0.02G_0$.
To illustrate the competition between quasiparticle tunneling and Andreev reflection,
we show in Fig.~\ref{fig-NS2}(d) the contribution of these two processes to the conductance
peak as a function of the normal state conductance. As expected, at very low transmissions
the peak height scales linearly with $G_{\rm N}$, which corresponds to the regime in which
single-quasiparticle tunneling dominates the subgap transport. In this regime, the Andreev
contribution scales as $G^2_{\rm N}$. Then, there is a crossover to a sublinear regime, which 
occurs when both contributions to the peak height are of the same order. Finally, for
$G_{\rm N} \gtrsim 0.02G_0$, the peak height is mainly determined by the Andreev reflection.
The sublinear behavior in this ``high-transmission" regime is a manifestation of the resonant
character of the Andreev reflection, or in other words, of the fact that the Andreev 
reflection is mediated by the presence of a sharp bound state. Notice also that in this
regime the single-quasiparticle contribution decreases upon increasing the transmission. 

\section{A superconducting tip: YSR states and multiple Andreev reflections} 
\label{sec-SS}

\begin{figure}[t]
\includegraphics[width=\columnwidth,clip]{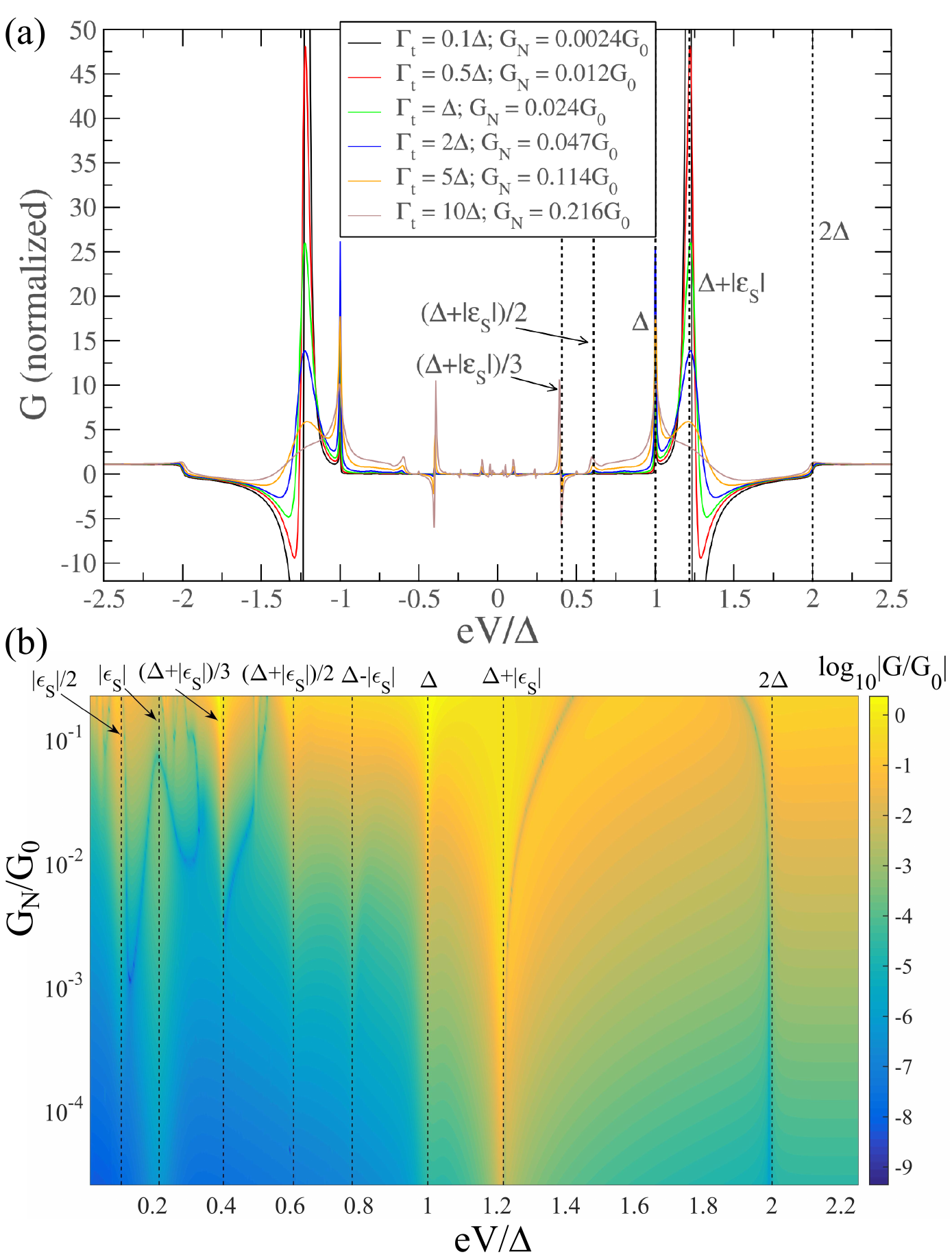}
\caption{(a) Differential conductance, normalized to its value at high bias, as 
a function of the bias voltage for the case of a superconducting tip. The different 
curves correspond to different values of the tunneling rate, as indicated in the 
legend where we also specify the value of the normal state conductance. The values 
of the different parameters of the model are: $\Gamma_{\rm S} = 100\Delta$, $J=80\Delta$, 
$U = 0$, $\eta_{\rm t} = 0.001\Delta$, $\eta_{\rm S} = 0.01\Delta$, and $k_{\rm B}T = 0.01\Delta$. 
The vertical dotted lines indicate the values of several relevant energies. Here,
$|\epsilon_{\rm S}|$ is the absolute energy of the YSR states as given by Eq.~(\ref{eq-YSR2}). 
(b) The logarithm (to the base 10) of the absolute value of the differential conductance 
$G=dI/dV$, normalized by the quantum of conductance $G_0=2e^2/h$, as a function of the bias 
voltage and the normal state conductance. The rest of the parameters have the same values as 
in panel (a). The vertical dashed lines indicate the values of different relevant energies, 
see upper part of the graph.}
\label{fig-SS1}
\end{figure}

In this section we shall discuss the current-voltage characteristics in the case in
which the tip is also superconducting, which is the central goal of this work. For 
simplicity, we shall assume that the tip and the substrate have the same gap that 
we shall denote as $\Delta$. In Fig.~\ref{fig-SS1}(a) we illustrate the rich
subgap structure that appears in the differential conductance as the transmission of
the junction increases in a case in which $J=80\Delta$ and $U=0$. The different curves
correspond to different values of the tunneling rate $\Gamma_{\rm t}$, while the coupling
to the substrate is kept constant and equal to $\Gamma_{\rm S} = 100\Delta$
(this value will be used throughout the whole section). We see new peaks appearing in the
differential conductance as the transmission increases, apart from the peaks at $eV =
\pm (\Delta + |\epsilon_{\rm S}|)$ that already appear in the deep tunnel regime due to 
the contribution of single-quasiparticle tunneling. Obviously, those additional features 
must originate from the contribution of various kinds of Andreev reflections, as we shall 
clarify below. To support our interpretation, we have added several vertical dashed lines 
at specific energies/voltages to the graph.  

To get further insight into the origin of the subgap structure, we present in Fig.~\ref{fig-SS1}(b)
a more systematic study of the evolution of the differential conductance as a function of 
the normal state conductance for the same case as in Fig.~\ref{fig-SS1}(a). Notice that for
convenience we are plotting here the absolute value of the conductance in logarithmic 
scale and we focus on positive voltages due to the electron-hole symmetry in this example.
We see that in the deep tunnel regime (for $G_\mathrm{N} < 10^{-3}G_0$), the conductance 
spectra are dominated by the presence of a peak at $eV=\Delta + |\epsilon_{\rm S}|$. This is 
the hallmark expected from the tunneling of single quasiparticles, as we discussed in section
\ref{sec-tunneling-spectra}, and it has been reported in numerous experimental studies
\cite{Heinrich2018}. There is also a feature at $eV = 2\Delta$ due to quasiparticle tunneling
connecting the gap edges of both electrodes, but it is much less pronounced due to the absence
of the BCS singularities in the LDOS of the magnetic impurity. It is worth mentioning that 
most experimental studies report a pronounced conductance peak at the sum of the gap energies 
of the tip and the substrate, something that cannot be explained with a model like ours in 
which there is a single current pathway through the impurity. It has been suggested that 
these pronounced coherent peaks can be explained by the presence of a second, non-magnetic 
channel that involves another orbital/level in the impurity \cite{Huang2019a}. Notably, there
are two additional peaks visible in this regime at $eV = |\epsilon_{\rm S}|$ and $eV = \Delta$,
although their heights are orders of magnitude smaller than the one at $eV=\Delta + 
|\epsilon_{\rm S}|$. One might be tempted to attribute these peaks to different kinds of 
Andreev reflections, but in fact those features can be accurately reproduced with the 
tunneling approximation of section \ref{sec-tunneling-spectra} and, therefore, they must 
originate from single-quasiparticle tunneling. The peak at $eV = |\epsilon_{\rm S}|$ can be
explained by the existence of a small but finite DOS inside the gap in the STM tip due to a
finite value of the broadening parameter $\eta_{\rm t}$ ($\eta_{\rm t} = 0.001\Delta$ in this
example). This peak simply occurs when the chemical potential of the tip is aligned with the
empty YSR state and we have checked that its height scales linearly with the normal state 
conductance $G_\mathrm{N}$. This conductance peak was reported in Ref.~\cite{Randeria2016} 
and it was correctly interpreted as a consequence of an ``imperfect" tip. On the other hand, 
the peak at $eV = \Delta$ is due to the finite DOS inside the gap region in the impurity
(induced by the substrate) due to the finite value of $\eta_{\rm S}$  ($\eta_{\rm S} = 
0.01\Delta$ in this example) and it appears when the gap edge of the tip is aligned to the
chemical potential of the impurity-substrate system. This conductance peak has also been
observed experimentally, e.g.\ in Ref.~\cite{Randeria2016}. More important for the discussion 
in this work is the fact that when the junction transmission (or normal state conductance)
increases, one observes the appearance of a whole zoo of conductance peaks, whose energies 
are identified in Fig.~\ref{fig-SS1}(b) with the help of vertical dashed lines. The main goal 
of the rest of this section is to understand the physical origin of those features. 

\begin{figure}[t]
\includegraphics[width=\columnwidth,clip]{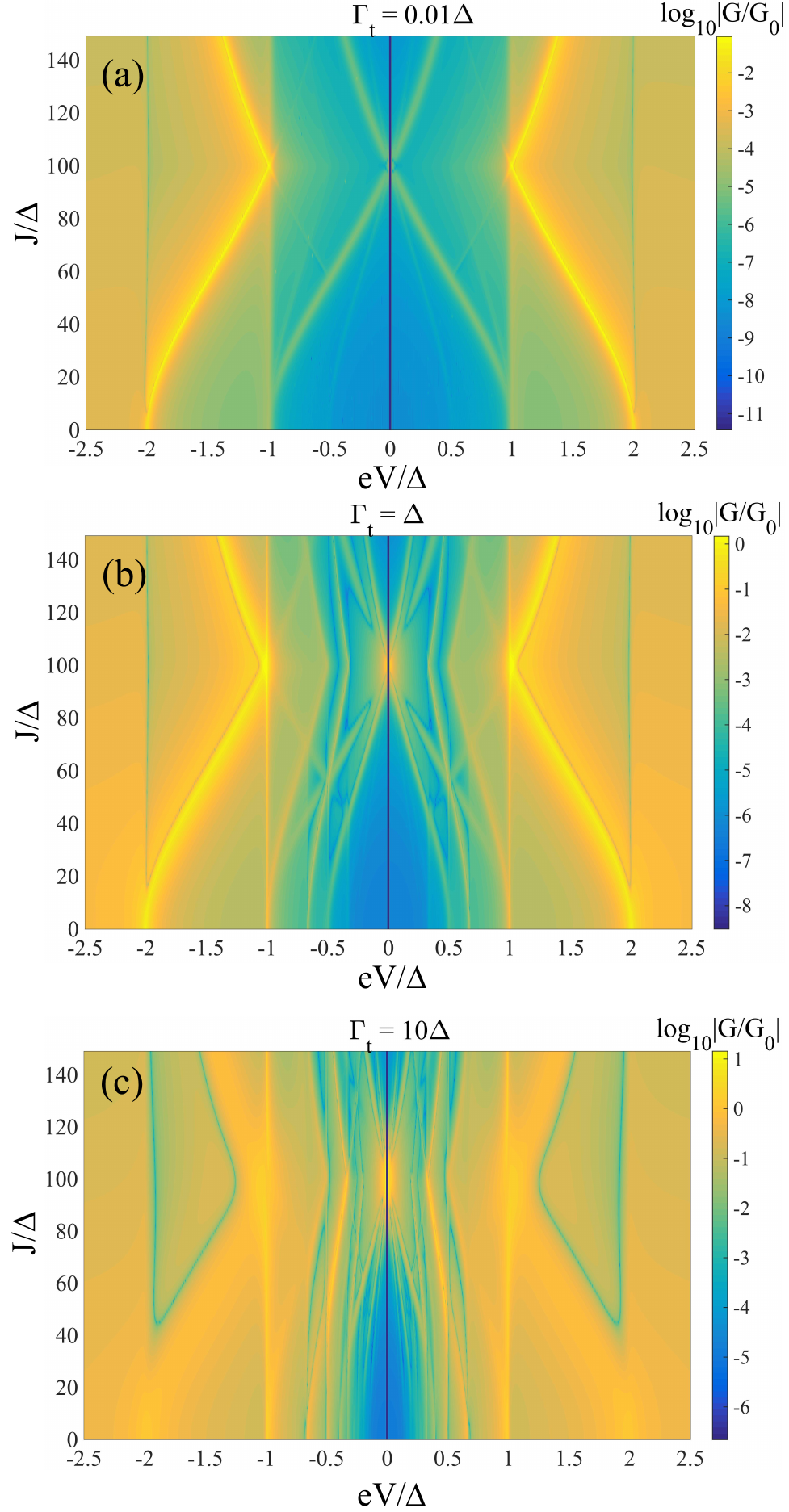}
\caption{Panels (a-c) show the logarithm (to the base 10) of the absolute value of the 
differential conductance $G=dI/dV$, normalized by the quantum of conductance $G_0=2e^2/h$, 
as a function of the bias voltage and the exchange energy for three different values of the 
tunneling rate: $\Gamma_{\rm t}/\Delta = 0.01, 1, 10$. The values of the different parameters 
are: $\Gamma_{\rm S} = 100\Delta$, $U = 0$, $\eta_{\rm t} = 0.001\Delta$, $\eta_{\rm S} = 
0.01\Delta$, and $k_{\rm B}T = 0.01\Delta$.}
\label{fig-SS2}
\end{figure}

An important hint on the origin of the different peaks in the subgap conductance can be obtained 
by analyzing how they shift when the energy of the YSR states is modified, for instance, by
changing the exchange energy. This is what we illustrate in Fig.~\ref{fig-SS2} where we
show the evolution of the differential conductance with the exchange energy for three
different values of the tunnel rate $\Gamma_{\rm t}$ and $U=0$. The case of $\Gamma_{\rm t} =
0.01\Delta$ in panel (a) corresponds to the tunnel regime and these results can be reproduced 
with the tunneling approximation of section \ref{sec-tunneling-spectra} (not shown here). As
expected, we see that the conductance spectra are largely dominated by the appearance of two
peaks that disperse with the energy of the YSR states as $eV=\pm (\Delta + |\epsilon_{\rm S}|)$. 
Other features that are also visible, albeit much less prominent, appear at $eV = 
\pm |\epsilon_{\rm S}|$, $eV = \pm 2\Delta$, and $eV = \pm \Delta$. As we explained in the 
previous paragraph, all of these features can be explained in terms of single-quasiparticle 
tunneling. For a higher value of the normal state conductance, like $\Gamma_{\rm t} = \Delta$ 
in panel (b), there appears a large variety of conductance peaks that cannot be explained by 
the tunneling of individual quasiparticles. Finally, when the normal state conductance is 
above $\sim 0.1G_0$, different conductance peaks start to overlap and the subgap structure 
becomes very complex, as we illustrate in Fig.~\ref{fig-SS2}(c) for $\Gamma_{\rm t} = 
10\Delta$.

\begin{figure}[t]
\includegraphics[width=\columnwidth,clip]{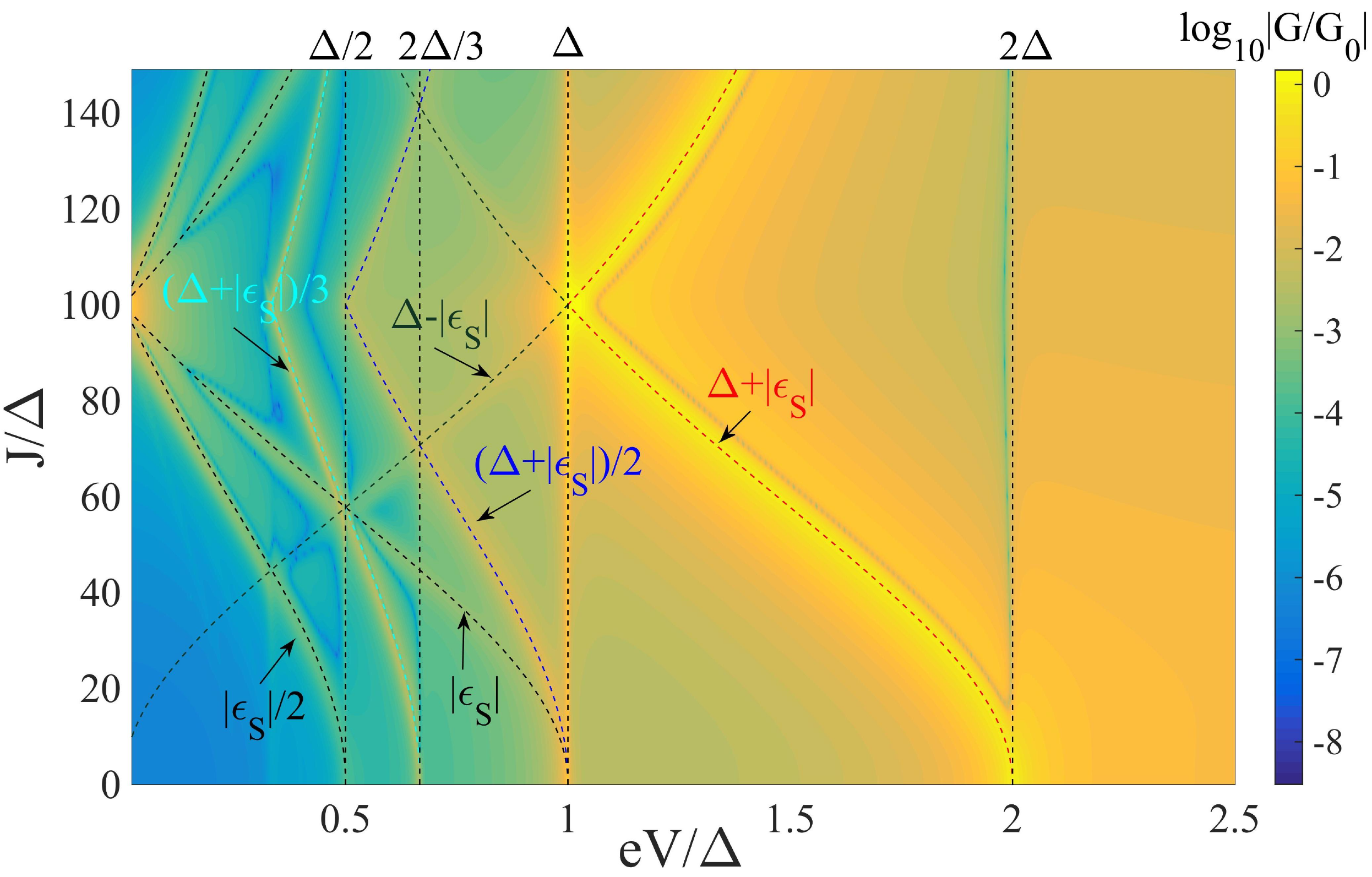}
\caption{The logarithm (to the base 10) of the absolute value of the differential conductance 
$G=dI/dV$, normalized by the quantum of conductance $G_0=2e^2/h$, as a function of the bias 
voltage and the exchange energy for $\Gamma_{\rm t} = \Delta$. The values of the other parameters 
are: $\Gamma_{\rm S} = 100\Delta$, $U = 0$, $\eta_{\rm t} = 0.001\Delta$, $\eta_{\rm S} = 0.01\Delta$, 
and $k_{\rm B}T = 0.01\Delta$. The different dashed lines correspond to various relevant energies,
as indicated in the legend, which describe the maxima of the subgap structure. The value of
$|\epsilon_{\rm S}|$ was taken from Eq.~(\ref{eq-YSR2}).}
\label{fig-SS3}
\end{figure}

To elucidate the relevant tunneling processes giving rise to the different features
of the subgap structure, it is important to identify the exact energies at which
these features appear. This is done in detail in Fig.~\ref{fig-SS3} for the case of 
$\Gamma_{\rm t} = \Delta$, panel (b) in Fig.~\ref{fig-SS2}, where we focus on positive 
voltages. We see that there are three different types of energies describing the
position of those features. First, there are energies, like $2\Delta/n$ with $n=1,2, \dots$, 
that only involve the gap energy. Second, there are energies that are a combination of the 
gap energy and the energy of the YSR bound states like, e.g., $\Delta \pm |\epsilon_{\rm S}|$ 
or $(\Delta + |\epsilon_{\rm S}|)/2$. Finally, there are energies like $|\epsilon_{\rm S}|$
and $|\epsilon_{\rm S}|/2$ that only involve the energy of the YSR states. With this 
identification we are now ready to propose the whole family of tunneling events that
can take place in our system.

\begin{figure*}[t]
\includegraphics[width=\textwidth,clip]{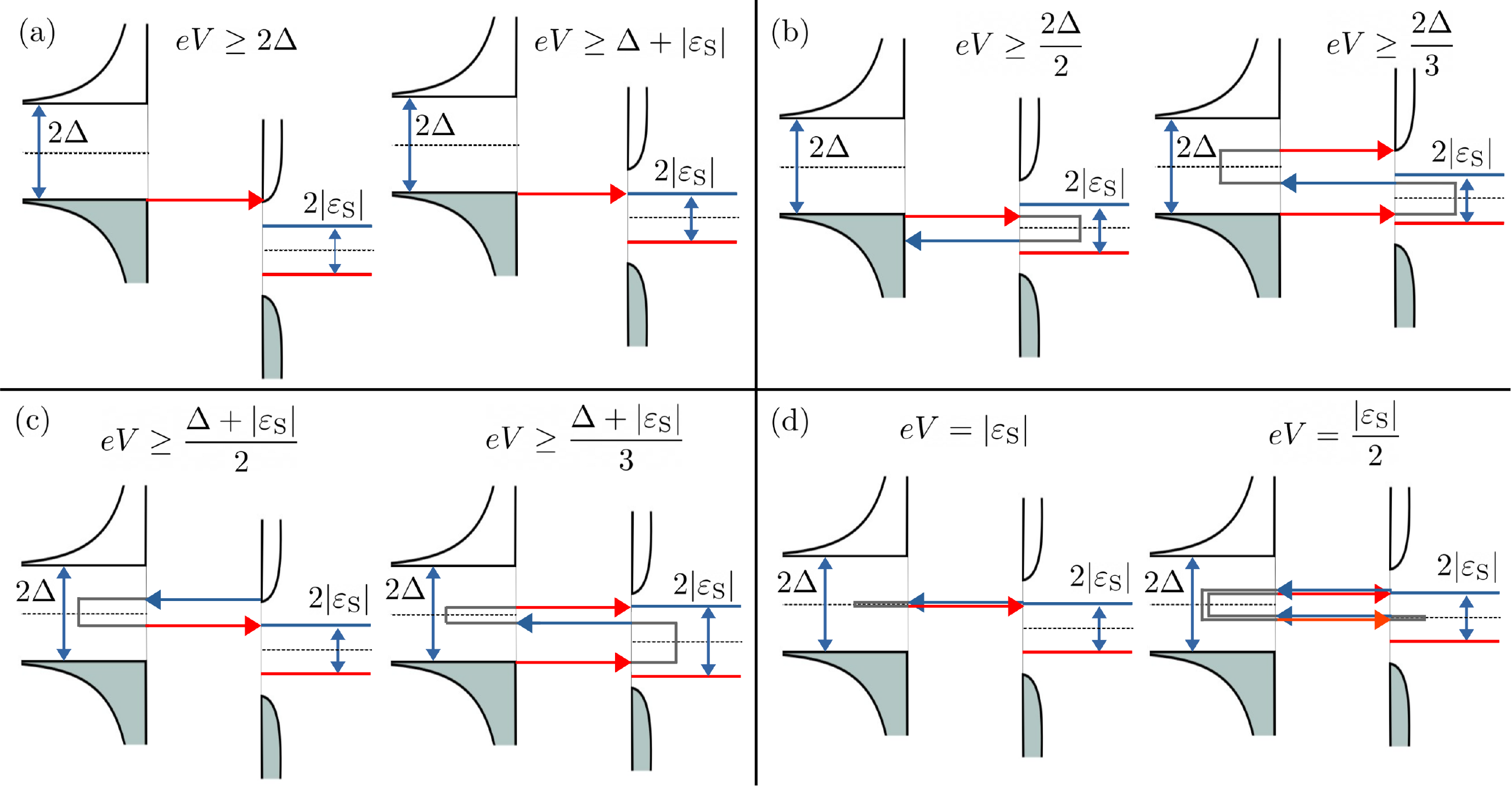}
\caption{Relevant tunneling processes giving rise to the subgap structure. Here, the 
left electrode is the superconducting tip and the right one is the impurity coupled to 
the substrate and their respective density of states are shifted by the bias voltage. 
The red lines correspond to electron-like quasiparticles and the blue ones 
to quasi-holes. In all cases, we indicate the threshold voltage at which they start to 
contribute to the current. (a) Single-quasiparticle processes that may (right) or may not 
(left) involve the YSR states. These are first-order processes in the tunneling rate or 
in the normal state conductance. (b) Standard MARs that do not involve any YSR state. 
They give rise to the subgap structure at $eV = \pm 2\Delta/n$ with $n > 1$, but they 
may also give resonant contributions at other values of the bias voltage (see main text). 
They are of order $n$ in the normal state conductance. (c) MARs that start or end in a 
YSR state. They give rise to the subgap structure at $eV = \pm (\Delta + |\epsilon_{\rm S}|)/n$ 
with $n>1$. They are of order $n$ in the normal state conductance. (d) MARs involving 
both YSR states. Energetically speaking, they would have threshold voltages given by 
$eV = \pm |\epsilon_{\rm S}|/n$ with $n \ge 1$, but they are forbidden due to the full
spin polarization of the YSR states.}
\label{fig-processes}
\end{figure*}

In Fig.~\ref{fig-processes} we summarize the relevant processes that give rise to the 
different features in the subgap conductance in our system, which we group into four
distinct families. The first family, illustrated in Fig.~\ref{fig-processes}(a), is 
formed by single-quasiparticle tunneling events in which a single electron/hole is 
transferred through junction. Within these processes, we can differentiate between 
tunneling events in which a quasiparticle can tunnel from the continuum of an electrode 
to the continuum of the other electrode, see left scheme in Fig.~\ref{fig-processes}(a), 
and events that start (end) in the continuum of states of tip and end (start) in the 
empty YSR state in the impurity, see right scheme in Fig.~\ref{fig-processes}(a). 
The first type is obviously responsible for the conductance feature at $eV = \pm 2\Delta$, 
while the second one gives rise to the conductance peaks at $eV = \pm (\Delta + 
|\epsilon_{\rm S}|)$. As discussed above, if there is some residual DOS inside the gap 
region in the electrodes, one can have additional single-quasiparticle tunneling events, 
which we shall not discuss here again. Moreover, if the temperature is not too low, 
there is another important single-quasiparticle process connecting the gap edge of the
tip with the lowest energy YSR state of the impurity that can be partially empty due to
the finite temperature. This well-known process gives rise to a conductance peak at 
$eV = \pm (\Delta - |\epsilon_{\rm S}|)$ that has been observed in numerous experiments,
see e.g.\ Ref.~\cite{Ruby2015} and references therein. Thus, one is tempted to explain 
the conductance peak at $eV = \Delta - |\epsilon_{\rm S}|$ in Fig.~\ref{fig-SS3} as the 
result of the tunneling of thermally excited quasiparticles. However, the temperature 
in that example is too low ($k_{\rm B}T = 0.01\Delta$) and we shall propose an alternative
explanation below (at the end of next paragraph). Let us conclude this discussion by saying 
that while the contribution of the single-quasiparticle processes is expected to be 
proportional to the normal state conductance, to leading order, the dependence on the junction
transmission might be more complicated in the case of the resonant processes involving 
the YSR states. As we explained in the previous section, this is actually the case when 
the transmission is sufficiently high such that the tunneling rate becomes larger 
than the natural broadening (inverse lifetime) of the bound states.  

The second family of tunneling processes are the standard MARs, which are schematically
represented in Fig.~\ref{fig-processes}(b). These are MARs that start and end in the 
continuum of states of the electrodes. They contribute to the subgap structure by 
increasing the conductance at their threshold voltages $eV = \pm 2\Delta/n$ with $n \ge 2$ 
(this is the usual subharmonic gap structure in the absence of magnetism) and in every 
process a charge equal to $ne$ is transferred. The absence of singularities in the LDOS of 
the impurity reduces the probability of the odd MARs like the one of order 3 shown on
the right hand side in Fig.~\ref{fig-processes}(b). More importantly, these MARs can
give resonant contributions, where their probability is greatly enhanced, when
during the cascade of reflections a quasiparticle hits the energy of a YSR state in
the impurity. Thus, for instance, the probability of the second-order Andreev reflection 
in Fig.~\ref{fig-processes}(b) is resonantly enhanced when $eV = \pm (\Delta +
|\epsilon_{\rm S}|)$. Notice that this is nothing else than the resonant Andreev
reflection that was discussed in the previous section for a normal-conducting tip. 
Therefore, this Andreev reflection competes with the single-quasiparticle process 
connecting the continuum of the tip DOS with the YSR state and it eventually dominates
the peak height at this bias when the junction transmission is sufficiently high.
This competition was nicely discussed in Ref.~\cite{Ruby2015}, both experimentally and
theoretically. In a similar way, other standard MARs can become resonant at certain
voltages. For instance, the third-order MAR in Fig.~\ref{fig-processes}(b) is resonantly
enhanced when $eV = \pm (\Delta - |\epsilon_{\rm S}|)$, while $eV \ge
2\Delta/3$. This implies that $|\epsilon_{\rm S}| \le \Delta/3$. At a first glance,
this resonant condition might explain the appearance of the conductance peak at $eV = 
\Delta - |\epsilon_{\rm S}|$ in Fig.~\ref{fig-SS3}. The alternative explanation of a 
peak due to thermally excited quasiparticles can be ruled out by the fact that such a 
peak would appear for any value of the YSR energy, which is not the case. The 
temperature in that example is simply too low for this quasiparticle process to give
a significant contribution. Notice, however, that a closer inspection of Fig.~\ref{fig-SS3}
shows that the peak at $eV = \Delta - |\epsilon_{\rm S}|$ extends up to $eV=\Delta/2$,
which suggests that another type of process is also at work in this case (see below). 

\begin{figure}[t]
\includegraphics[width=\columnwidth,clip]{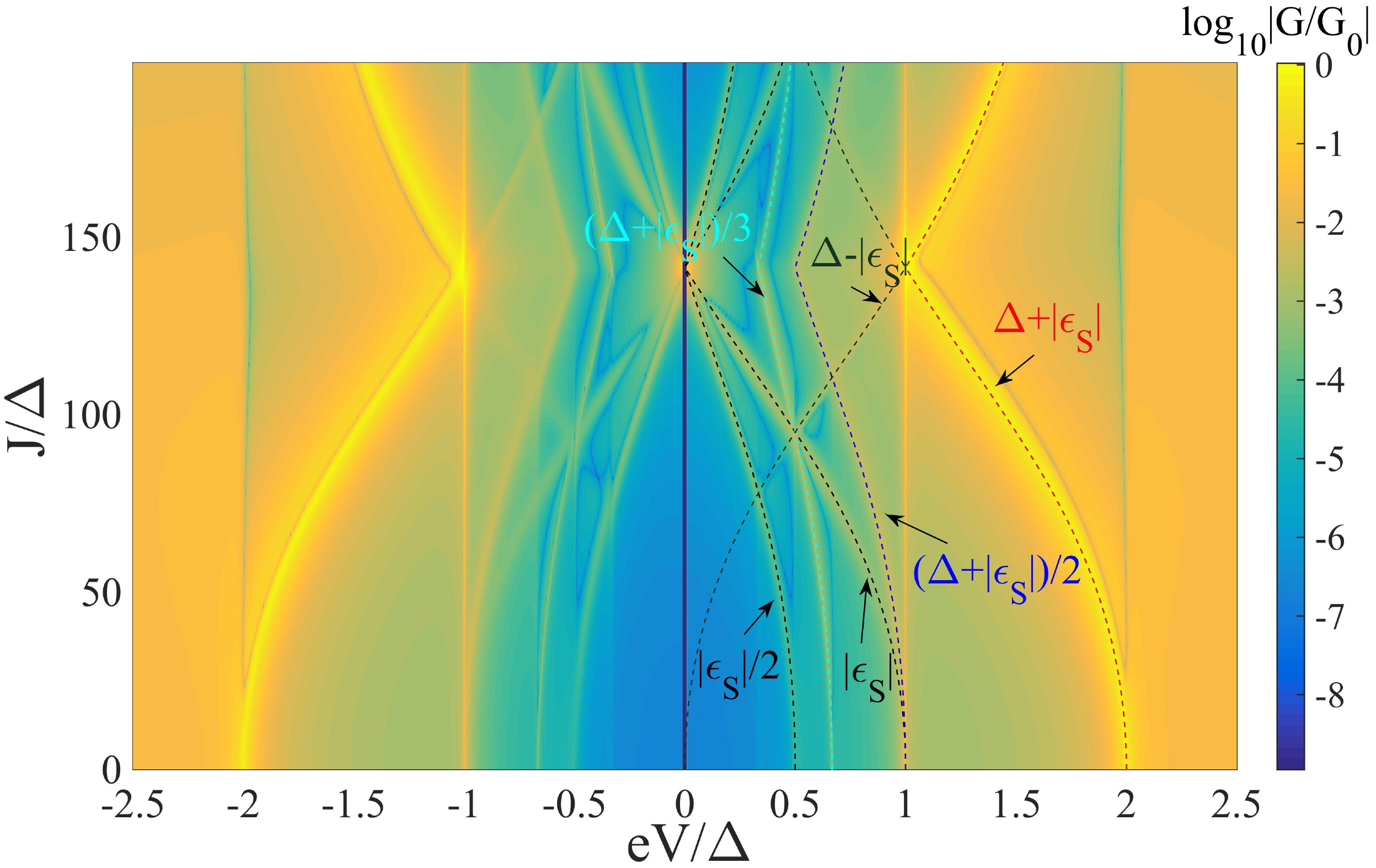}
\caption{The same as in Fig.~\ref{fig-SS3}, but for $U = 100\Delta$. Notice that in this
case the differential conductance depends on the bias polarity. The energy of the YSR
states, $|\epsilon_{\rm S}|$, for the dashed lines in the graph was taken from 
Eq.~(\ref{eq-YSR1}).}
\label{fig-SS4}
\end{figure}

Another family of Andreev reflections are those described in Fig.~\ref{fig-processes}(c)
in which a MAR either starts or ends in a YSR state. They give rise to the subgap structure 
at their threshold voltages $eV = \pm (\Delta + |\epsilon_{\rm S}|)/n$ with $n>1$ and they 
involve the transfer of $n$ charges across the junction. By comparing these processes with 
the standard MARs, we can see that, qualitatively speaking, the presence of the YSR states 
on the second superconducting contact reduces its ``effective gap" such that quasiparticle
states are present at $|\epsilon_\mathrm{S}|$ instead of $\Delta$. On the other hand, since
their probability critically depends on the DOS associated to a single YSR state, their
contribution to the differential conductance depends on the bias polarity (see below 
discussion of Fig.~\ref{fig-SS4}). Moreover, there is 
a basic difference between even and odd MARs of this kind. Since the odd ones must connect
the continuum of states of the tip, which does exhibit a BCS singularity, and a YSR state in
the impurity, they can give rise to a negative differential conductance (NDC), which is 
actually what we see in Fig.~\ref{fig-SS3}. Signatures of the occurrence of these MARs have
been reported experimentally in Refs.~\cite{Randeria2016,Farinacci2018}. Let us also say
that if there is some residual DOS inside the gap, for instance in the tip, one can also
have MARs connecting the YSR states with that finite in-gap DOS. Thus, for instance, it
is easy to convince oneself that it is possible to have such a second-order Andreev 
reflection connecting the lower YSR state and the residual DOS in the tip. This process
has a threshold (positive) voltage $eV = \Delta - |\epsilon_{\rm S}|$ and it is only
possible as long as $eV > |\epsilon_{\rm S}|$, which altogether implies that 
$|\epsilon_{\rm S}|< \Delta/2$. We think that this is indeed the process that gives 
the main contribution to the conductance peak at $eV = \Delta - |\epsilon_{\rm S}|$ in 
Fig.~\ref{fig-SS3}. In particular, this nicely explains why this peak is only visible for
voltages $eV \ge \Delta/2$.

Finally, we want to discuss a more exotic type of MAR that, in principle, could also 
exist in the presence of bound states. These MARs would start and end in a YSR bound
state, see Fig.~\ref{fig-processes}(d). Energetically speaking, these processes could
occur at voltages $eV = \pm |\epsilon_{\rm S}|/n$ with $n \ge 1$ and they would involve 
the transfer of a charge equal to $2ne$. Obviously, if they existed, their contribution 
would drastically depend on the bound state broadening and, given their resonant nature,
they should give rise to a NDC. Given their hypothetical threshold voltages, it is tempting 
to assign the features that we see in Fig.~\ref{fig-SS3} at $eV = |\epsilon_{\rm S}|$ and 
$eV = |\epsilon_{\rm S}|/2$ to the occurrence of these resonant MARs. However, a closer 
inspection of the diagrams of Fig.~\ref{fig-processes}(d), and in particular of the spin 
of the quasiparticles involved in these processes, shows that these MARs require a bound 
state to have a finite DOS of both spin species, which is not the case for YSR states. 
So, in other words, these MARs are strictly forbidden due to the full spin polarization 
of the YSR bound states \cite{note2}. Then, the remaining question concerns the origin 
of the subgap structure at the energy of the YSR states and subharmonics of it. As discussed
above, we think that the peak at $eV = |\epsilon_{\rm S}|$ in Fig.~\ref{fig-SS3}, which does
not exhibit an NDC, is mainly due to a single-quasiparticle process. On the other hand, we 
attribute the peak at $eV = |\epsilon_{\rm S}|/2$ in Fig.~\ref{fig-SS3} to the contribution 
of a second-order Andreev reflection connecting a YSR state and the residual DOS in the 
gap region on the impurity site. Such a process has precisely a threshold voltage 
$eV = |\epsilon_{\rm S}|/2$. Let us say that we are not aware of any experimental 
observation of this latter conductance feature. 

From our discussions above, we have concluded that the contributions to the differential
conductance attributed to tunneling processes which either start or end in
a single YSR state can depend on the bias polarity. To illustrate this fact, we show
in Fig.~\ref{fig-SS4} the results for the differential conductance in a case similar to 
that of Fig.~\ref{fig-SS3}, but in which the electron-hole is broken ($U=100\Delta$). 
These results not only confirm our statement above, but they also show that the same
series of conductance peaks appears in the subgap conductance when there is no electron-hole 
symmetry in the system.

\section{Discussion and conclusions} \label{sec-conclusions}

There are a number of ways in which this work could be extended. First of all, it 
would be desirable to extend the full counting statistics (FCS) theory of MARs to the
case of magnetic impurities discussed in this work \cite{Cuevas2003,Cuevas2004}. The
FCS technique allows us to unambiguously classify the tunneling processes according 
to the charge transferred. This could shed some additional light on the nature of the
different transport processes in the complex situation investigated here. On the other 
hand, in many experiments several pairs of YSR states are reported. In this sense,
it would be nice to study the interplay between different bound states and how this
is reflected in the current-voltage characteristics for arbitrary transparency. 
In principle, this could be done within the framework of the Anderson model used in 
this work by including additional energy levels or orbitals in the impurity. Another
interesting extension could be the analysis of the role of dynamical Coulomb blockade
in our system. It is known that this dynamical effect, which results from the interaction
of tunneling electrons with the electromagnetic environment, ultimately limits the
energy resolution of tunneling experiments and it may have an important impact,
especially, at very low temperatures \cite{Ast2016}. It would also be
of great interest to study the role of electron correlations (beyond the mean
field approximation used here) in all the transport properties discussed in this
work. However, this is very challenging and it continues to be an important open 
problem \cite{Andersen2011}. Finally, our theory is ideally suited to describe the
tunneling between magnetic impurities exhibiting their respective YSR states, which
has been experimentally reported for the first time very recently \cite{Huang2019b}.
This is a problem that we shall tackle in a forthcoming paper.

So, to conclude, we have presented in this work a microscopic theory of the quantum 
transport through individual magnetic impurities coupled to superconductors. Motivated
by recent STM-based experiments, we have studied the interplay between YSR states 
and (multiple) Andreev reflections in these systems with the help of a combination of
a mean-field Anderson model and nonequilibrium Green's function techniques. We have 
been able to identify the different tunneling processes and, in particular, we have
predicted the occurrence of a large variety of Andreev reflections mediated by 
YSR states. Moreover, we have provided very precise guidelines on how to identify
the contribution of these processes in actual experiments. From a more general 
perspective, our work provides further insight into spin-dependent Andreev transport
that can also be of interest for the community of superconducting spintronics
\cite{Linder2015}.

\acknowledgments

The authors would like to thank Alfredo Levy Yeyati, Jakob Senkpiel, Robert Drost, 
Joachim Ankerhold, Ciprian Padurariu, and Bj\"orn Kubala for insightful discussions.
A.V.\ and J.C.C.\ acknowledge funding from the Spanish Ministry of Economy and
Competitiveness (MINECO) (contract No.\ FIS2017-84057-P). This work was funded 
in part by the ERC Consolidator Grant AbsoluteSpin (Grant No.\ 681164) and by 
the Center for Integrated Quantum Science and Technology (IQ$^\textrm{\small ST}$).
R.L.K., W.B., and G.R.\ acknowledge support by the DFG through SFB 767 and Grant 
No.\ RA 2810/1. J.C.C.\ also acknowledges support via the Mercator Program of 
the DFG in the frame of the SFB 767.

$^{\dagger}$These authors contributed equally to this work.


\begin{thebibliography}{00}
\bibitem{note1}
The origin of this model is the following. Mean field calculations show that the spin
degeneracy can be broken in presence of local Coulomb interaction for two (bare) spin
degenerate levels. In the mean-field theory, the parameter $U$ and $J$ are calculated 
self-consistently. In our approach, we analyze the model in the parameter space of $U$ 
and $J$.
\end{thebibliography}
\end{document}